\pdfoutput=1
\documentclass[useAMS,usenatbib]{mn2e} 
\usepackage{lscape}
\usepackage{graphicx} 
\usepackage{amssymb} 
\usepackage[usenames]{color}

\title[{\it Kepler} observations in Upper Scorpius]{{\it Kepler} observations of A-F pre-main sequence stars in Upper Scorpius: Discovery of six new $\delta$~Scuti and one $\gamma$~Doradus stars} 
\author[V. Ripepi et al.]{V. Ripepi$^{1}$\thanks{E-mail: ripepi@oacn.inaf.it}, L. Balona$^{2}$, G. Catanzaro$^{3}$, M. Marconi$^{1}$,
F. Palla$^{4}$, M. Giarrusso$^{5}$ 
\\  
$^{1}$INAF-Osservatorio Astronomico di Capodimonte, Via Moiariello 16, I-80131, Napoli, Italy\\
$^{2}$South African Astronomical Observatory, PO Box 9, Observatory 7935, Cape Town, South Africa\\
$^{3}$INAF-Osservatorio Astrofisico di Catania, Via S.Sofia 78, I-95123, Catania, Italy\\ 
$^{4}$Osservatorio Astrofisico di Arcetri, Firenze, Italy\\
$^{5}$Universit\`a degli studi di Catania, Via S.Sofia 78, I-95123 Catania, Italy, Italy\\
} 
 
\date{Accepted   Received ; in original form } 
 
\pagerange{\pageref{firstpage}--\pageref{lastpage}} \pubyear{2002} 
 
\def\LaTeX{L\kern-.36em\raise.3ex\hbox{a}\kern-.15em 
 T\kern-.1667em\lower.7ex\hbox{E}\kern-.125emX} 
 
\begin{document} 
 
\label{firstpage} 
 
\maketitle 
 
\begin{abstract} 
We present light curves and periodograms for 27 stars in the young Upper
Scorpius association (age=$11 \pm 1$\,Myr) obtained with the Kepler
spacecraft.  This association is only the second stellar grouping to
host several pulsating pre-main sequence (PMS) stars which have been observed 
from space.  

From an analysis of the periodograms, we identify six
$\delta$~Scuti variables and one $\gamma$~Doradus star.  These are most 
likely PMS stars or else very close to the zero-age main sequence. Four of 
the $\delta$~Scuti variables were observed in short-cadence mode, which
allows us to resolve the entire frequency spectrum.  For these four stars, we 
are able to infer some qualitative information concerning their ages.  For 
the remaining two $\delta$~Scuti stars, only long-cadence data  are available,
which means that some of the frequencies are likely to be aliases.  
One of the stars appears to be a rotational variable in a hierarchical triple 
system. This is a particularly important object, as it allows the possibility 
of an accurate mass determination when radial velocity observations become
available.  We also report on new high-resolution echelle spectra obtained
for some of the stars of our sample.

\end{abstract} 
 

\setcounter{equation}{0}

\section{Introduction}

A complete understanding of the star formation process requires the
ability to predict how the properties of young stars depend on their
initial conditions. Despite significant efforts, a key
observational problem remains: the reliable determination of
the masses and ages of the youngest pre-main-sequence (PMS) stars once
they become optically visible, after the main accretion phase.
Stellar masses are needed to investigate the shape of
the initial mass function (IMF) and its possible dependence on the
properties of the parent cloud.  However, obtaining reliable stellar masses
is a very difficult task. The only direct method is to use eclipsing 
binary systems where individual masses of the components can be obtained
with great accuracy from the radial velocity and light curve. However, 
eclipsing systems for PMS stars are rare.  Another commonly adopted procedure is 
to locate the stars in the Hertzpung-Russell (HR) diagram, using 
estimates of their effective temperatures and luminosities.  Masses can
be obtained by comparing their position in the HR diagram with
theoretical evolutionary tracks, assuming the chemical composition
\citep{Hillenbrand2004}. 

Recently, considerable advances have been made in our understanding of 
PMS evolution \citep{Palla1999,Siess2000,Baraffe2002,Tognelli2011}.  
However, differences in the evolutionary tracks remain due to the different
treatment of convection, different opacities and differences in the 
zero-point of the calculated ages.  This results in considerable
differences in mass estimates of single stars from their effective
temperatures and luminosities.  Without adequate observational constraints, 
it is impossible to determine the correct assumptions in the evolutionary model
calculations.  For this reason, it is very important to
obtain independent determinations of PMS masses and ages.

In this context, intermediate-mass PMS stars (with mass in the range
1--8 M$_{\odot}$, also called Herbig Ae/Be stars) are particularly
useful.  After the seminal work by \citet{Marconi1998}, who established
the locus of the theoretical instability strip in the HR diagram, it
has become clear that these stars pulsate as $\delta$~Sct variables
(see e.g. \citealt{Ripepi2002,Ripepi2003,Marconi2004,Ripepi2006a,Ripepi2006b,Zwintz2008}
and references therein).  More recently, high-precision photometry from 
space missions such as {\it MOST} and {\it CoRoT} showed that
intermediate-mass PMS stars can also pulsate as hybrid
$\delta$~Sct/$\gamma$~Dor stars \citep{Ripepi2011} or pure $\gamma$~Dor
stars \citep{Zwintz2013}.  This is interesting because the occurrence of
low frequencies in $\delta$~Sct stars (i.e. hybrid pulsation) was 
originally thought to be exceptional.  However, {\it Kepler} observations
have shown that all $\delta$~Sct stars have low frequencies
\citep{Balona2014}.  The reason why the hybrid concept arose is due to the
fact that the low frequencies attain sufficient amplitudes to be visible
from the ground only in a narrow temperature range.  The mechanism which
drives low frequencies in hot $\delta$~Sct stars is currently unknown.

The additional information provided by pulsation can be used to
  construct detailed models of the varying internal structure of these
  stars as they contract toward the main sequence, in order to compare
  the predicted frequencies to the observed ones
  \citep{Ruoppo2007,Dicriscienzo2008,Casey2013}. 
 The physics of PMS stellar evolutionary models can
  therefore be tested.  Several efforts have recently been made to
  analyze the properties of known PMS pulsators \citep[see
  e.g.][]{Zwintz2014}.  However, the analysis is hampered by several
  factors such as the uneven quality of the observational data (ground
  and space time series data have very different properties), the lack
  of precise spectroscopically derived stellar parameters, the
  possibly different intrinsic properties of cluster and field PMS
  pulsators, and difficulties in the treatment of rapidly rotating stars.

In this context, PMS pulsators belonging to clusters or associations are of
particular interest, since the member stars share the same chemical
composition, age and interstellar absorption, greatly 
simplifying the comparison between theory and observations.  This is the 
case, for example, of NGC\,2264, a well-studied cluster where 9 PMS pulsators 
have been observed from space \citep{Zwintz2014}.  The Upper Scorpius 
association (USco) is perhaps of even greater interest due to the large number 
of known A--F members \citep{deZeeuw1999,Pecaut2012} and the availability of 
accurate individual distances based on {\it Hipparcos} parallaxes.
Moreover, stellar parameters have been estimated for most of the member
stars.

USco is part of the larger Scorpius-Centaurus (Sco-Cen) association, a young 
($\sim$5--10 Myr) and relatively close ($\sim$120--150\,pc) region of
recent star formation \citep{deBruijne1999,deZeeuw1999} that includes the  
Upper Centaurus-Lupus (UCL) and Lower Centaurus-Crux (LCC).  USco is the 
youngest of them and contains many intermediate-mass members, 
making it an ideal target to investigate the 
whole extension of the PMS instability strip. Indeed, its member stars are 
sufficiently bright ($\sim 7<V<10$\,mag) to acquire extremely precise  
photometry both from the ground and from space.

Due to the loss of a second spacecraft reaction wheel, the {\it Kepler} 
satellite ended data collection in the original Cygnus field after four 
years of continuous monitoring. By pointing near the ecliptic plane, the 
{\it Kepler} spacecraft is able to minimize pointing drift, acquiring data with 
much reduced photometric precision.  This mode of operation is called the K2
project.\footnote{See http://keplerscience.arc.nasa.gov/K2/ for  
details.}  The USco region falls partially within the field of view of the {\it
Kepler} satellite during the {\it K2-C02} campaign, becoming a primary target 
for our purposes. In this work we present the results of a search for pulsating 
PMS stars among the USco member stars observed by {\it Kepler}.

The structure of the paper is the following: in Section 2 we describe
the {\it Kepler} data and the techniques of data reduction and analysis; in Section
3 we present the classification of the variables; Section 4 reports a
the results of the spectroscopic follow up for a sample of interesting
templates; Section 5 includes a discussion about the evolutionary
status of the target stars; Section 6 reports a discussion of the
properties of the pulsating stars. Finally, a brief summary closes the paper.

\section{{\it Kepler} observations}

{\it Kepler} observations usually consist of 30-min exposures (long cadence
or LC mode), but a few stars can be observed with 1-min exposures (short
cadence or SC mode).  Observations of 16 PMS stars in USco, 7 of which were
also observed in SC mode, were obtained by {\it Kepler}.  The {\it K2-C02} campaign 
started on 2014 Aug 23 and ended on 2014 Nov 10, for a total duration of about 
78.7\,d.  The selected stars come from the lists of \citet{deBruijne1999} and 
\citet{Pecaut2012}.  The spectral types of the selected stars range
from A2 to F7 (to embrace the entire $\delta$ Sct and $\gamma$ Dor
instability strips, taking also into account possible errors on the
spectral type values by 1-2 subclasses or more). We also 
searched the {\it Kepler K2} data archive at MAST (Mikulski Archive for Space 
Telescopes)  for additional A--F members of USco and found an additional 9
objects (2 in SC mode), mainly A0 stars, giving a total of 27 USco
members.  The targets and their main properties are listed in Table~\ref{tabGen}.

\begin{table*} 
\scriptsize 
\begin{center}
\caption{Stellar Parameters For A-type Upper Sco Member. The meaning
  of the different columns is the following: (1) EPIC number; (2) HIP
  identifiers; (3) mode of observation: LC and SC stands for Long and Short
  cadences, respectively; (4) variable classification; (5) {\it
    Kepler} magnitude; (6) Spectral Type; (7) Interstellar absorption;
(8) parallax; (9) logarithm of the effective temperature; (10)
logarithm of the luminosity in solar units. All the data reported in
columns (6-10) are from \citet{Pecaut2012}; (11) Disk classification according to \citet{Luhman2012}: d/e=debris/evolved transitional disk; f=full disk. }  
\label{tabGen} 
\begin{tabular}{ccccccccccc} 
\hline 
\hline 
\noalign{\medskip} 
EPIC  &  HIP & Mode & Var. Type & KepMag  &  ST & A$_V$ & $\pi$ &
$\log T_{\rm eff}$ &  $\log L/{\rm  L}_\odot$ & Disk\\ 
\noalign{\medskip} 
     &  &  &  & mag      &  & mag &mas&    &    \\ 
(1)   &  (2)    & (3)         & (4)         &    (5)&(6) &(7)&(8) &(9)&(10) &(11)\\ 
\noalign{\medskip} 
\hline 
\noalign{\smallskip}                                                 
202842502  &   79733   &  LC          &   --      &    9.054  &     A1mA9-F2  &  1.25  $\pm$   0.04  &  4.51  $\pm$ 0.43  & 3.964 $\pm$  0.019 & 1.48    $\pm$ 0.092     &  -                                                    \\
202876718  &   79878   &  LC          &   --        &    7.183  &     A0V            &   0.00  $\pm$   0.02  &  7.35  $\pm$ 0.55 &  3.980 $\pm$  0.031 & 1.407  $\pm$ 0.07     &d/e                                                       \\
203120347  &   80586 &    LC          &   --        &  8.211   &   F5IV-V & 0.00$\pm$0.10 &  7.02$\pm$0.65 & 3.814$\pm$0.011 & 0.93$\pm$0.09                                   &-                    \\
203399155  &   80311  &   LC         &    --       &    8.878  &     A1V            &   0.93  $\pm$   0.05  &  5.51  $\pm$ 0.49 &  3.964 $\pm$  0.019 & 1.266  $\pm$ 0.08      &-                                                        \\
203660895  &   79097 &    LC          &   --        &  9.155   &   F4V              & 0.48$\pm$0.12 & 6.99$\pm$0.57 & 3.822$\pm$0.009 & 0.90$\pm$0.09                            &-                \\
203712541  &   78494 &    LC          &   --        &  7.722   &   A2mA7-F2 & 0.75 $\pm$ 0.06 & 7.26 $\pm$ 0.57 & 3.943 $\pm$ 0.016 & 1.395 $\pm$ 0.077                 &  -                               \\
203774126  &   80799 &    LC          &   --        &  7.900   &   A3V      & 0.25 $\pm$ 0.04 & 7.97 $\pm$ 0.62 & 3.932 $\pm$ 0.012 & 1.078 $\pm$ 0.068                        & d/e                              \\
203931628  &   80196   &  LC          &  DSCT      &    9.030  &     A1Vn           &  2.36  $\pm$   0.18   & 5.94  $\pm$ 0.56  & 3.964   $\pm$ 0.019 & 1.602  $\pm$ 0.102  &  d/e                                                        \\
204054556  &   79054   &  LC          &  GDOR      &    9.215  &     F3V            &    0.35  $\pm$   0.05  &  6.24 $\pm$  0.53 &  3.827 $\pm$   0.005&  0.78    $\pm$ 0.08   & d/e                                                        \\
204076987  &   79643 &    LC          &    --        &  9.393   &   F3V & 0.56$\pm$0.05 & 6.63$\pm$0.62 & 3.827$\pm$0.005 & 0.67$\pm$0.08                                          & d/e                \\
204175508  &   77960 &    LC          &  DSCT      &  8.410   &   A4IV/V   & 0.75 $\pm$ 0.02 & 8.56 $\pm$ 0.71 & 3.918 $\pm$ 0.012 & 0.973 $\pm$ 0.093                       &  -                              \\
204222666  &   78099   &  LC          &   --        &    7.790  &     A0V            &   0.59  $\pm$   0.03  &  7.42 $\pm$  0.59 &  3.980 $\pm$  0.031 & 1.367  $\pm$ 0.073      &d/e                                                      \\
204239132  &   80238   &  LC          &   --        &    7.848  &     A2.5V          &  0.74  $\pm$   0.09   & 8.00  $\pm$ 0.68  & 3.937  $\pm$ 0.014 & 1.389  $\pm$ 0.081    & d/e                                                        \\
204242194  &   79250   &  LC          &   --        &    7.829  &     A3III/IV       &  0.33  $\pm$   0.05   & 9.60  $\pm$ 0.75  & 3.932 $\pm$  0.012 & 0.963  $\pm$ 0.072    &d/e                                                        \\
204372172  &   80088 &    LC/SC      &  DSCT      &  9.000   &   A9V      & 0.61 $\pm$ 0.03 & 6.11 $\pm$ 0.66 & 3.872 $\pm$ 0.010 & 0.960 $\pm$ 0.095                      &   d/e                            \\
204399980  &   79476 &    LC/SC      &  DSCT/LPV       &  8.822   &   A8IVe    & 1.02 $\pm$ 0.40 & 7.56 $\pm$ 0.61 & 3.875 $\pm$ 0.010 & 1.006 $\pm$ 0.175              &      f                                   \\
204492384  &   79644 &    LC          &  ROT       &  10.024  &   F6V & 0.53$\pm$0.11 & 6.30$\pm$0.64 & 3.802$\pm$0.007 & 0.44$\pm$0.10                                        &-               \\
204494885  &   80130 &    LC/SC      &  DSCT      &  8.727   &   A9V      & 0.66 $\pm$ 0.02 & 6.41 $\pm$ 0.57 & 3.872 $\pm$ 0.010 & 1.089 $\pm$ 0.077                       &   -                             \\
204506777  &   78977 &    LC/SC      &  EA/ROT    &  8.752   &   F8V & 0.38$\pm$0.08 & 7.53$\pm$0.64 & 3.788$\pm$0.007 & 0.83$\pm$0.08                                        & -              \\
204638251  &   80059 &    LC/SC      &  DSCT      &  8.745   &   A7III/IV & 0.56 $\pm$ 0.07 & 7.74 $\pm$ 0.64 & 3.892 $\pm$ 0.014 & 0.907 $\pm$ 0.075                         & -                             \\
204662993  &   79977 &    LC/SC      &   --        &  9.099   &   F3V & 0.36$\pm$0.05 & 7.79$\pm$0.66 & 3.827$\pm$0.005 & 0.63$\pm$0.08                                      & d/e              \\
204810792  &   79083 &    LC          &   --        &  8.226   &   F3V & 0.76$\pm$0.13 & 7.16$\pm$0.63 & 3.827$\pm$0.005 & 1.13$\pm$0.09                                       &-                \\
204926239  &   79606 &    LC/SC      &   --        &  9.039   &   F8V & 0.81$\pm$0.14 & 9.19$\pm$0.79 & 3.788$\pm$0.007 & 0.65$\pm$0.09                                       & -               \\
204966512  &   80019   &  LC          &   --        &    8.396  &     A0V            &   1.03  $\pm$   0.07  &  7.81  $\pm$ 0.68 &  3.980 $\pm$  0.031 & 1.374  $\pm$ 0.084   &d/e                                                        \\
205002311  &   82218 &    LC          &  ROT?      &  9.008   &   F3V & 0.30$\pm$0.02 & 8.39$\pm$0.72 & 3.827$\pm$0.005 & 0.55$\pm$0.08                                       &d/e              \\
205089268  &   79156   &  LC          &   --        &    8.804  &     A0V            &   0.60  $\pm$   0.07  &  6.43  $\pm$ 0.54 &  3.980 $\pm$  0.031 & 1.366  $\pm$ 0.081  & d/e                                                       \\
205181377  &   79124   &  LC          &   --        &    8.630  &     A0V            &   0.78  $\pm$   0.05  &  6.51  $\pm$ 0.53 &  3.980 $\pm$  0.031 & 1.55   $\pm$ 0.078   & -                                                        \\
\noalign{\smallskip} 
\hline 
\end{tabular}
\end{center}
\end{table*}

\begin{figure*}                                       
\includegraphics[width=16cm]{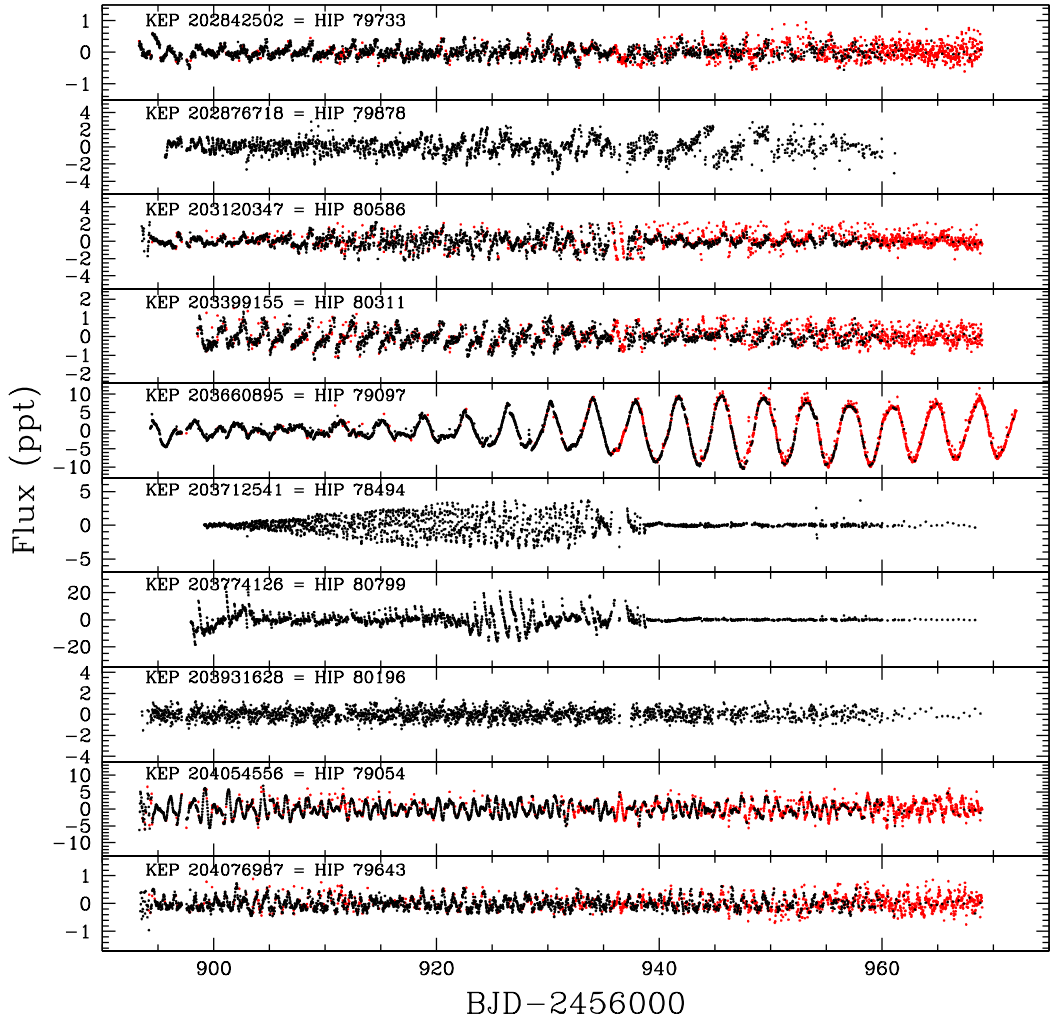}     
\caption{LC light curves for the 27 objects investigated in this 
  paper. Black and red points show data with quality flag equal or 
  greater than zero, respectively. For some objects the non zero flag 
  data was omitted because it was too scant.} 
\label{fig1} 
\end{figure*} 

\begin{figure*}                                     
\centering
\includegraphics[width=16cm]{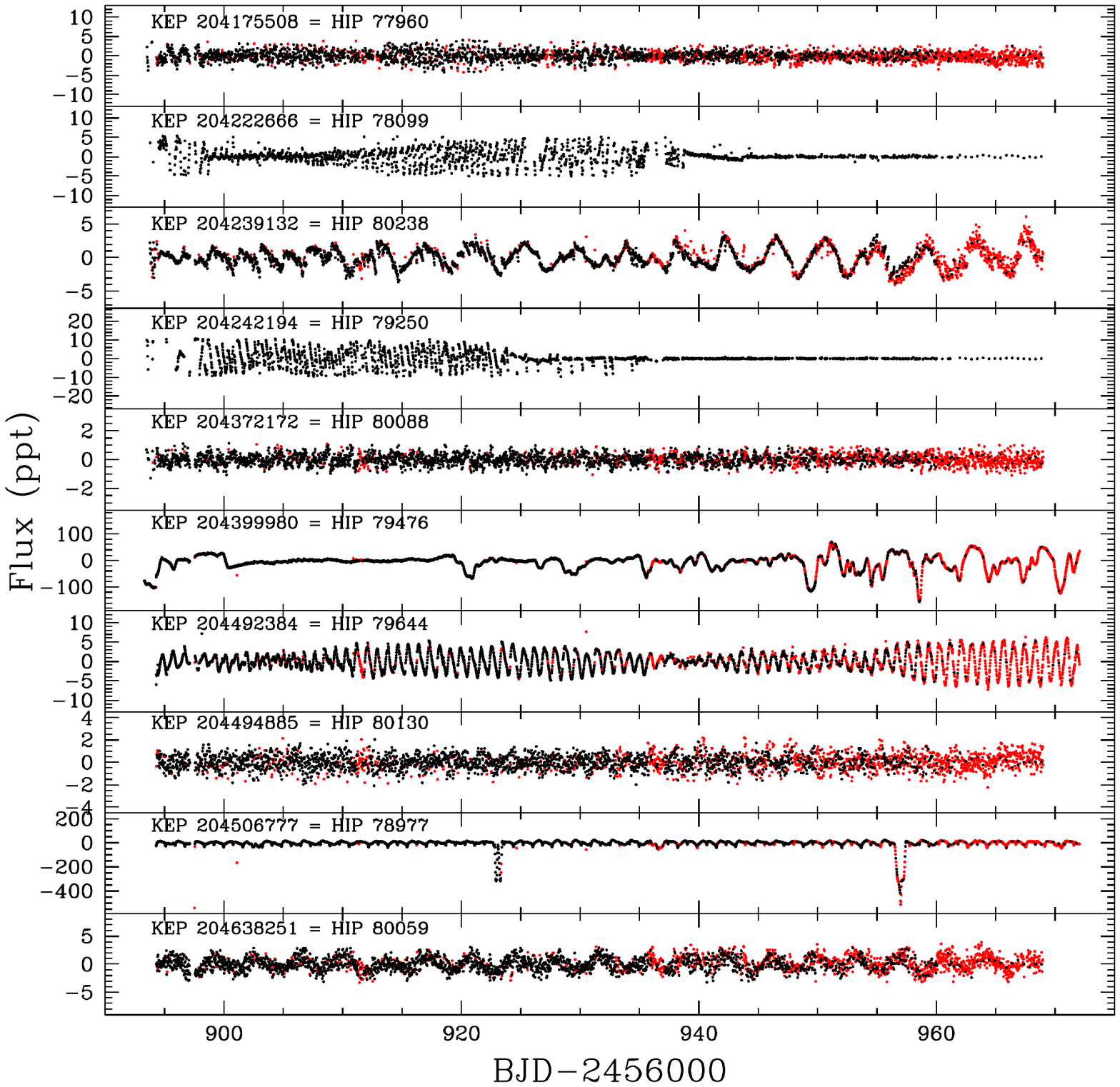}     
\contcaption{}
\end{figure*} 

\begin{figure*}                                       
\centering
\includegraphics[width=16cm]{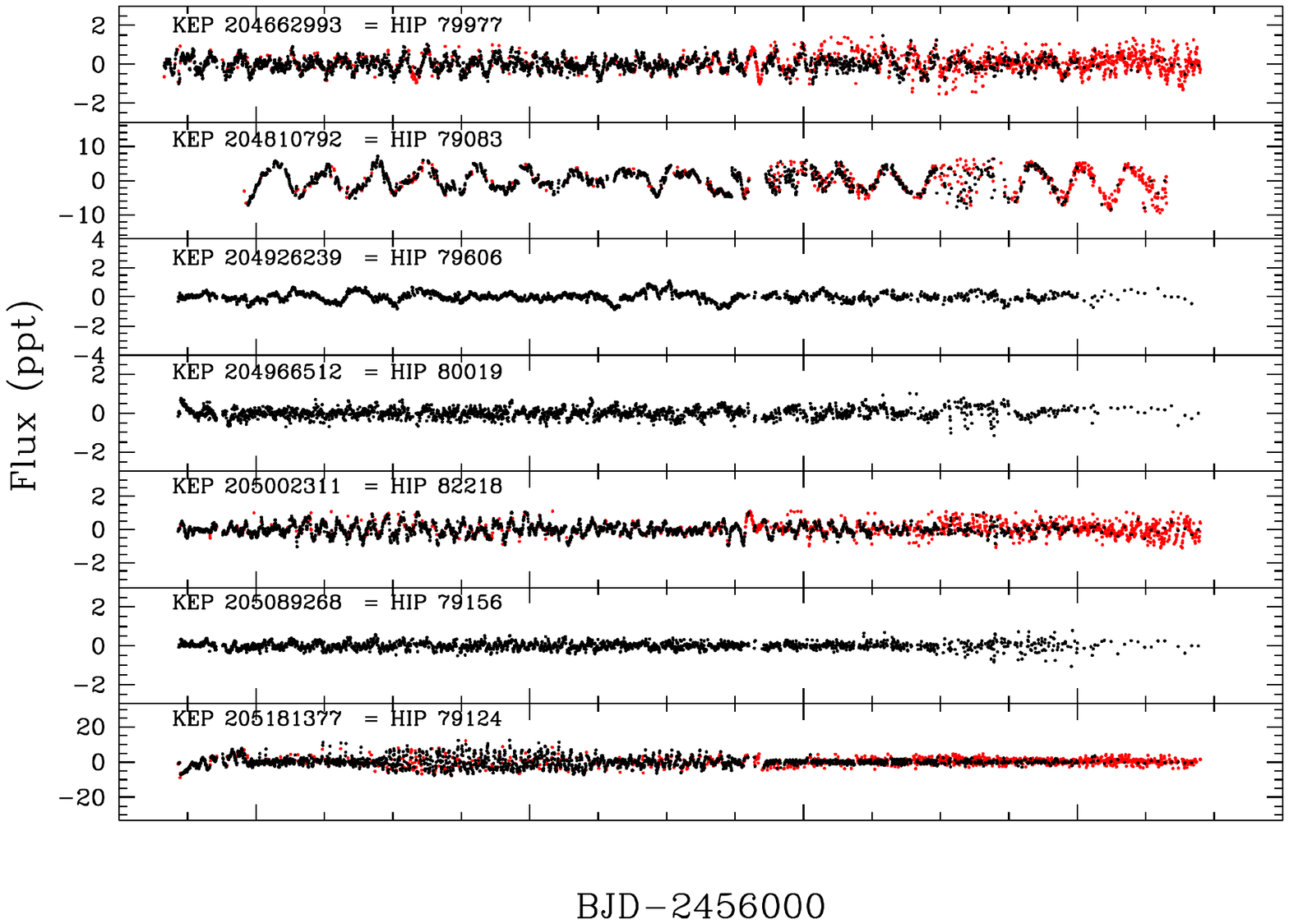}     
\contcaption{}
\end{figure*}

\begin{figure*}                                       
\includegraphics[width=16cm]{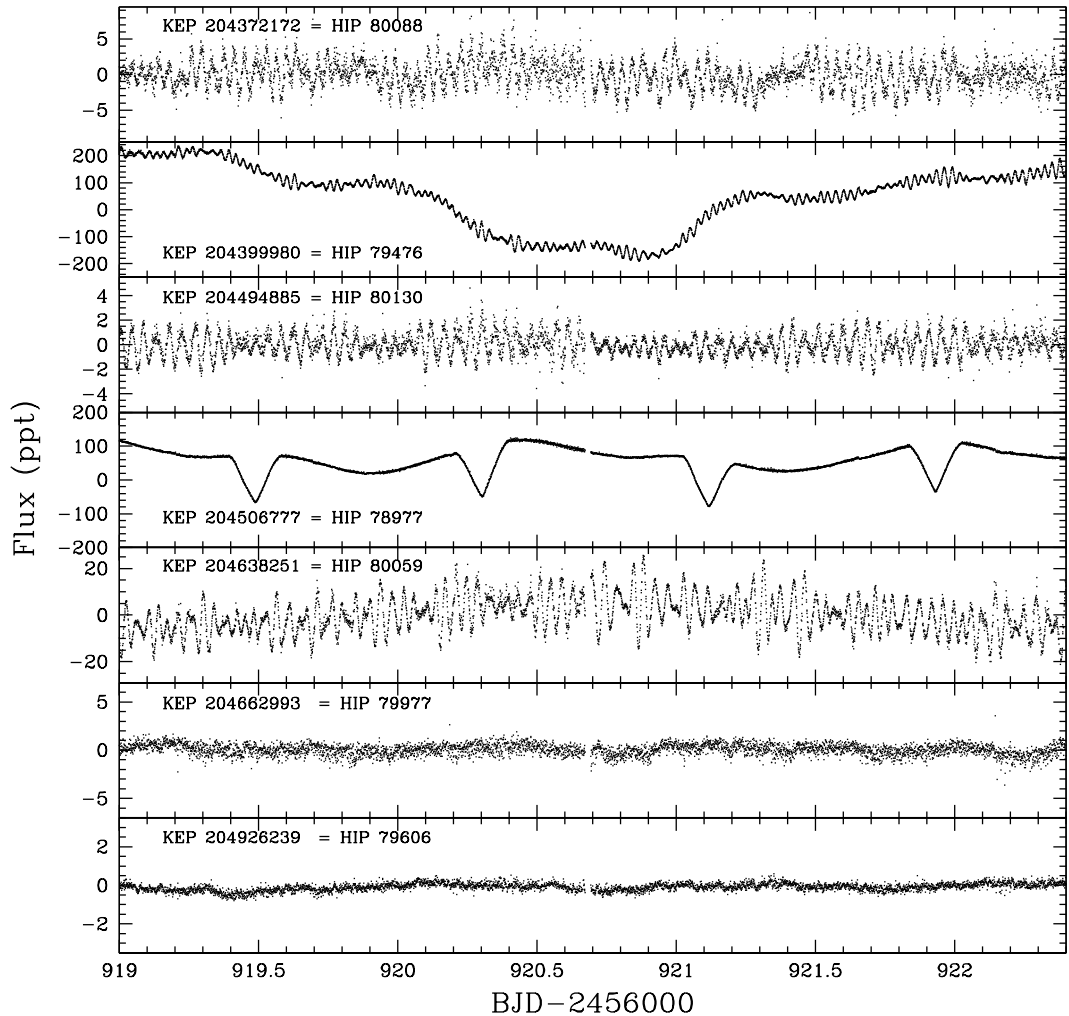}     
\caption{The figure shows selected portion of SC light curves for the seven 
targets observed in this cadence.}
\label{fig2} 
\end{figure*} 

\begin{figure*}                                       
\includegraphics[width=16cm]{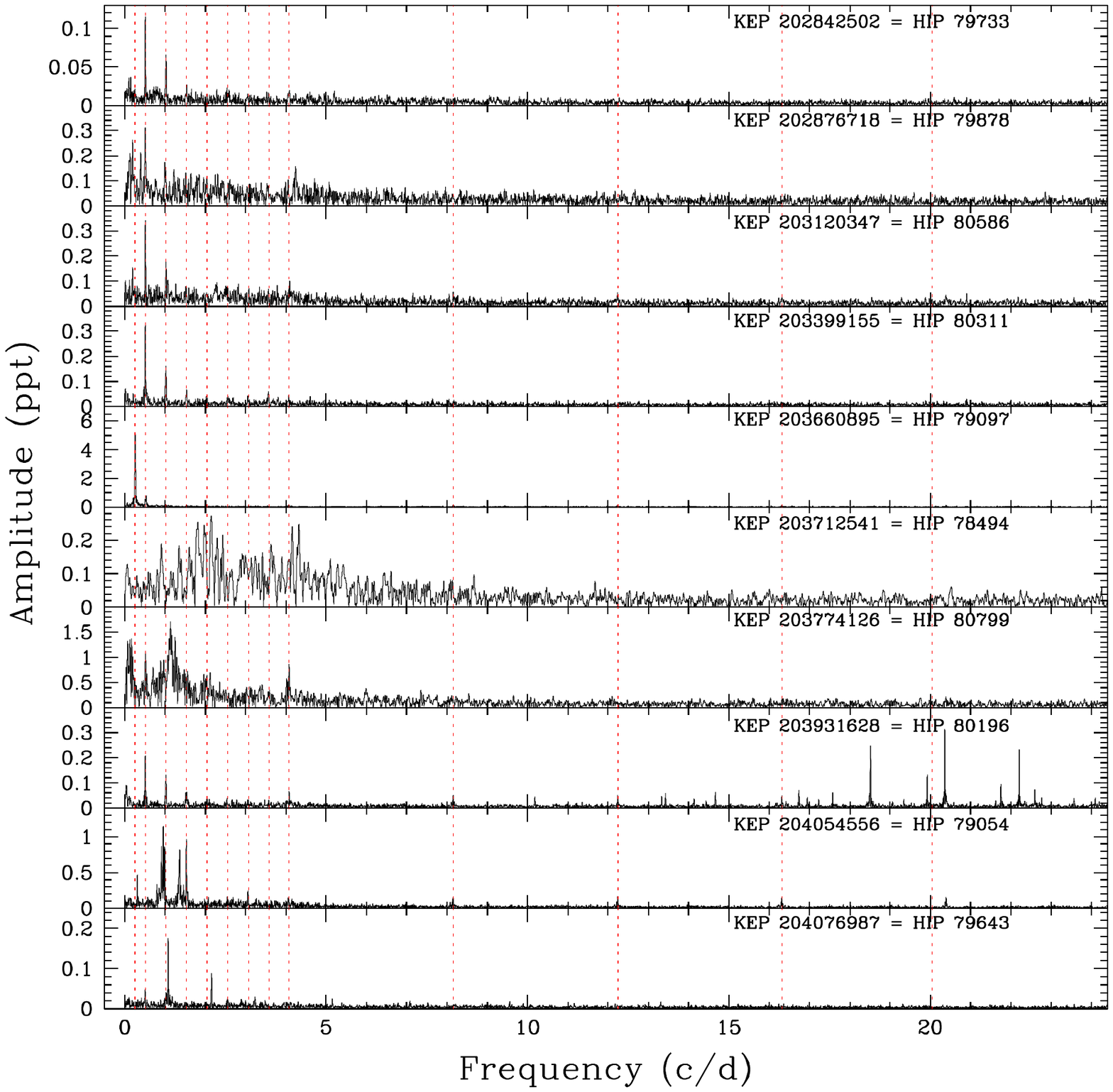}     
\caption{Fourier transform for the 27 objects with LC data. The dashed 
red lines show frequencies related to the satellite motion.} 
\label{fig3} 
\end{figure*} 

\begin{figure*}                                       
\includegraphics[width=16cm]{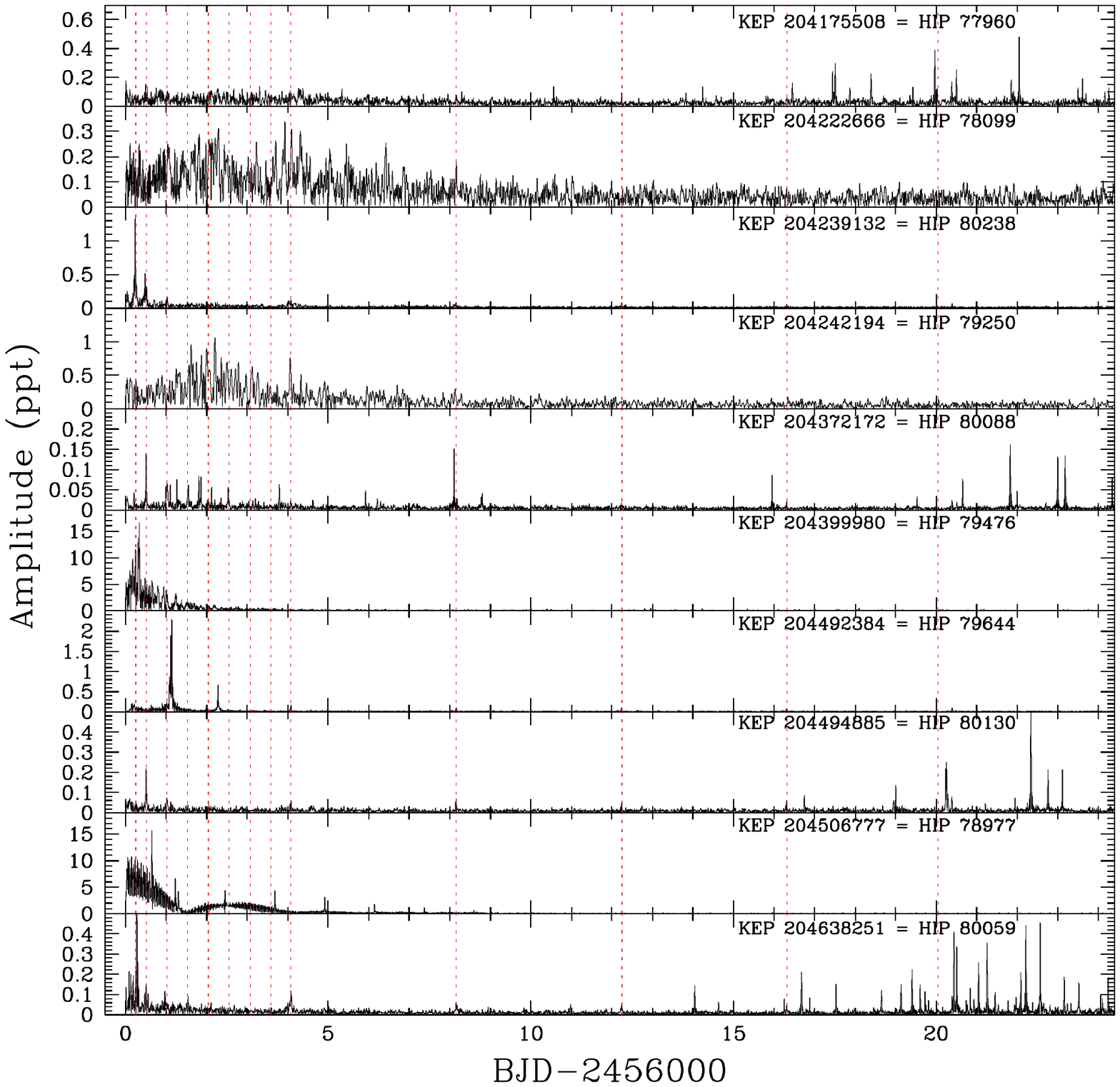}     
\contcaption{}
\end{figure*} 

\begin{figure*}                                       
\includegraphics[width=16cm]{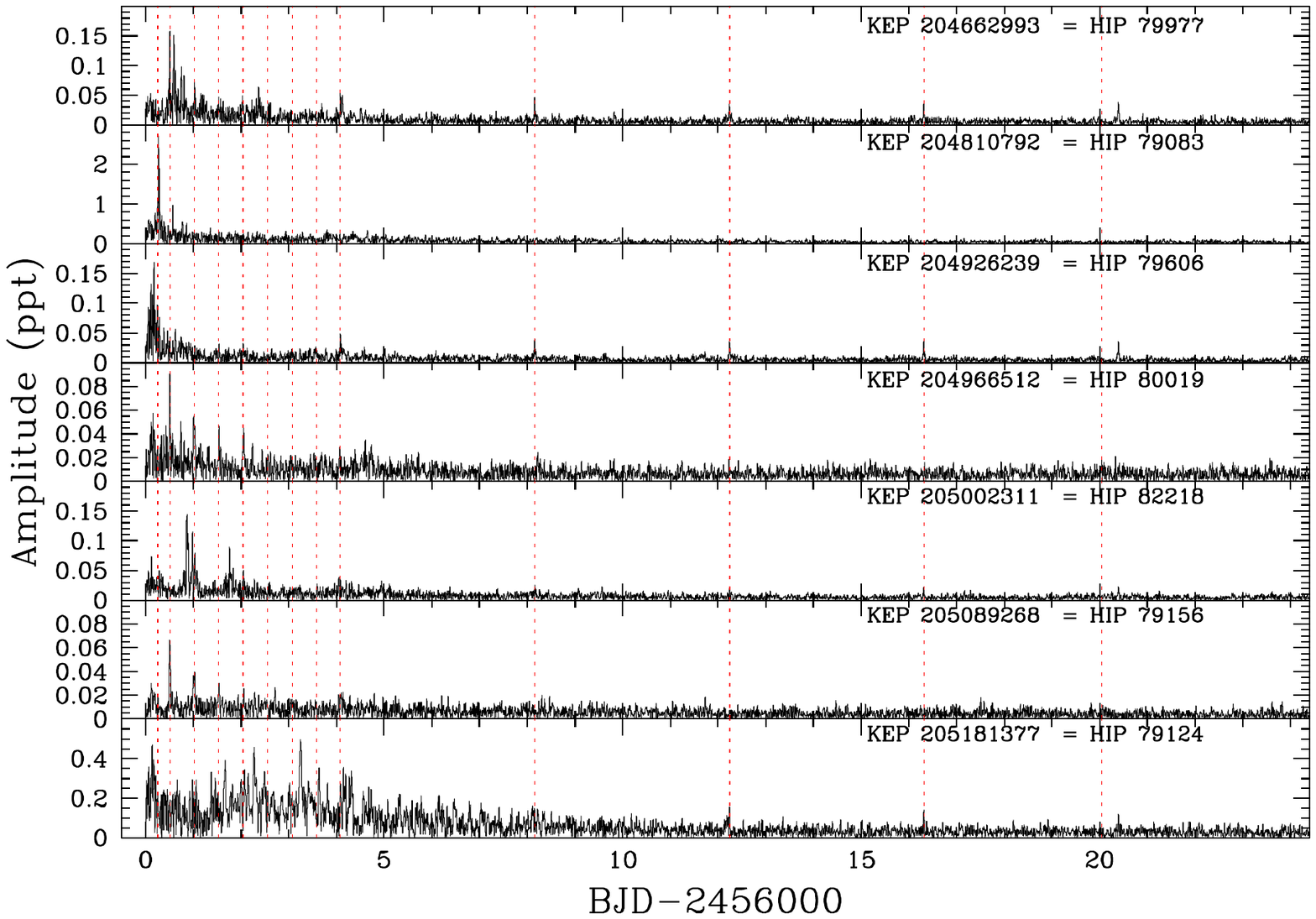}     
\contcaption{}
\end{figure*}

\begin{figure*}                                       
\includegraphics[width=16cm]{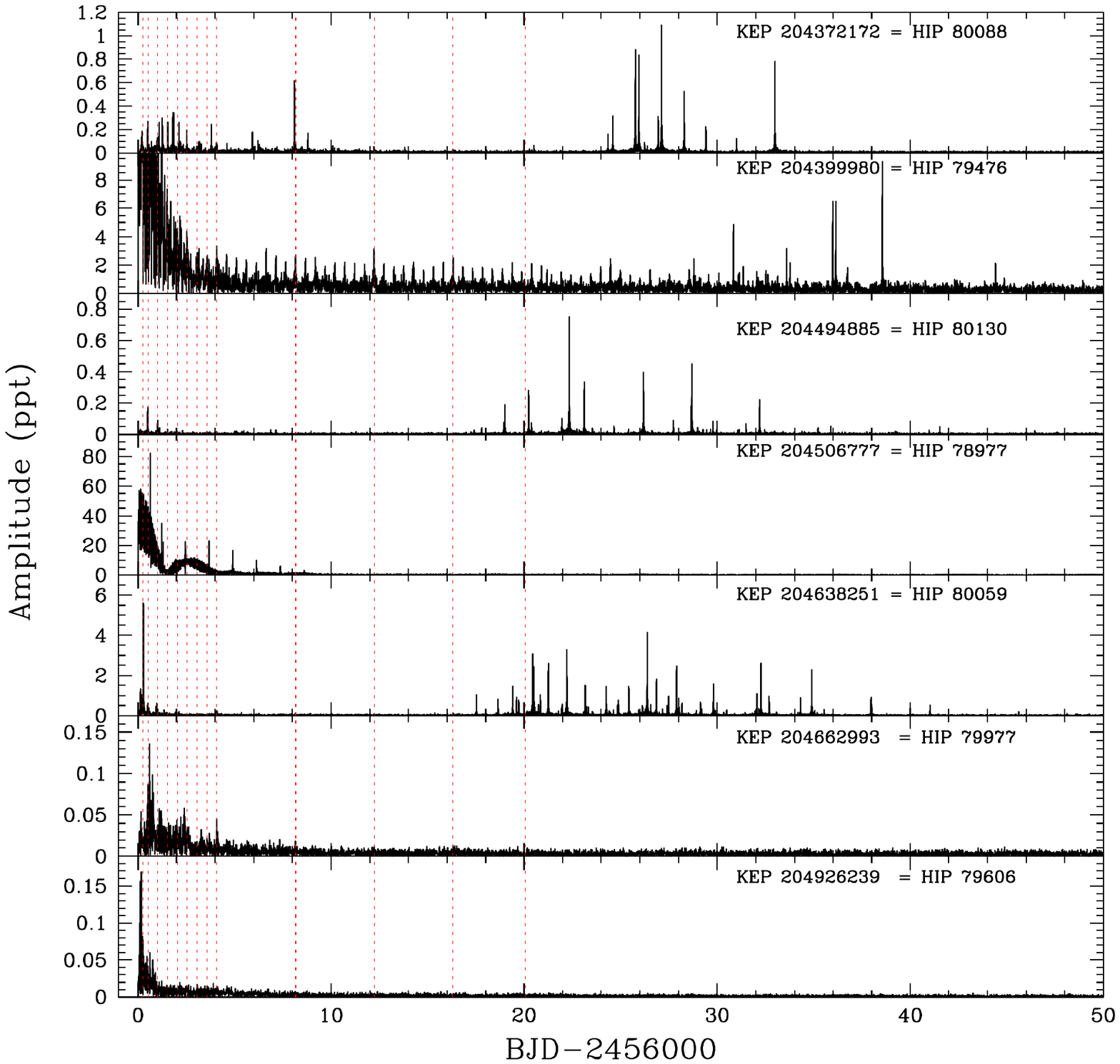}     
\caption{Fourier transform for the 7 objects with SC data. The dashed 
red lines show frequencies related to the satellite motion.} 
\label{fig4} 
\end{figure*}

\subsection{Data reduction and frequency analysis}

Data reduction was carried out by means of the  {\sc PyKe} software
\citep{Still2012} and also by our own custom software.  Simple aperture 
photometry (SAP) was extracted from raw pixel files and then corrected 
for the instrumental signatures present in the data. Spacecraft drift is 
the main component leading to increased photometric scatter compared to the 
original {\it Kepler} field.  Using SAP, the brightness of a target changes 
slowly as the star moves from its original location on the CCD until the 
thrusters are applied to bring the star back to its original location.  The 
periodogram therefore shows strong peaks at about 4.08\,d$^{-1}$ and its 
harmonics, which corresponds to the frequency at which the thrusters are 
applied.  It is possible to obtain a greatly improved light curve simply by 
locating these peaks in the periodogram and fitting and removing a truncated 
Fourier series.  Extraction of the light curve from the FITS files using SAP, 
fitting and removing the truncated Fourier series was done using custom
software.
  
A more sophisticated approach to correct the light curve was proposed 
by \citet{Vanderburg2014}.  In this technique the non-uniform pixel response 
function of the {\it Kepler} detectors is determined by correlating flux 
measurements with the spacecraft's pointing and removing the dependence. 
This leads to an improvement over raw SAP photometry by factors of 2--5, with 
noise properties qualitatively similar to {\it Kepler} targets at the same 
magnitudes.  There is evidence that the improvement in photometric precision 
depends on each target's position in the {\it Kepler} field of view, with worst 
precision near the edges of the field. Overall, this technique restores the 
median-attainable photometric precision within a factor of two of the original
Kepler photometric precision for targets ranging from 10--15 magnitude 
in the {\it Kepler} bandpass.  This technique is implemented in the {\sc
kepsff} task within the {\sc PYKE} software suite.  This is the method used in
our analysis.  Correction of the periodogram of the raw SAP photometry as
mentioned above is very useful as a check on possible artefacts which may
arise in the {\sc kepsff} task. 

The light curves for the 27 targets with LC data are shown in Fig.~\ref{fig1}, 
whereas Fig.~\ref{fig2} displays an enlargement of the light curves for the 7 stars 
with SC data.  Periodograms and frequency extraction were performed using the 
{\sc  Period04} software \citep{Lenz2005}.  The periodograms for these stars are 
shown in Fig.~\ref{fig3} and \ref{fig4}.

Problems in the low frequency regime may be expected given the image drift. It is 
therefore important to identify common frequencies in the periodograms which could 
be of instrumental origin or artefacts of the reduction process.  Most of the 
periodograms obtained from applying the \citet{Vanderburg2014} technique have a
peak at frequency $\nu = 0.51$\,d$^{-1}$ which can be identified as an
instrumental frequency.  In many stars the peak at half this frequency, $\sim$0.26\,d$^{-1}$, 
is also present.  Stars having this feature include EPIC\,202842502, 203660895, 
203931628, 204076987, 204494885, 204638251, 204810792 and 205089268.  Other artificial 
peaks are present and these must be recognized in order not to be misled in
classifying the star as a variable. In Fig.~\ref{fig3} and ~\ref{fig4}, we have 
shown with dashed lines the frequencies that are likely of instrumental
origin.  These are the harmonics of $\nu = 0.51$\,d$^{-1}$ and of 4.08\,d$^{-1}$.

\section{Classification of variables}
\label{Classification}

As an essential first step, it is necessary to examine the periodograms and
light curves of all stars in order to identify pulsating variables and other
interesting objects.  The only pulsating stars in the observed spectral
range that can be safely identified are $\delta$~Sct (characterized by the presence of one or more high
frequencies, i.e. frequencies in excess of about 5\,d$^{-1}$). 
$\gamma$~Doradus stars, which have multiple frequencies lower than about
5\,d$^{-1}$, are difficult to identify with certainty.  This arises because
of the possibility of spurious instrumental frequencies in this range as
well as irregular variability caused by circumstellar material which might
be present in these young stars.  Furthermore, one cannot ignore the
possibility that many stars may have spots which have different frequencies
due to differential rotation or finite lifetimes.  Bearing these difficulties
in mind, our attempted classifications are presented in Table~\ref{tabGen}. 

Five stars are definitely $\delta$~Sct (DSCT) variables: EPIC\,203931628, 
204175508, 204372172, 204494885, and 204638251.  
At a first look EPIC\,204399980 seems to be an irregular long-period variable.  
However, a closer inspection of the SC light curve and its periodogram (Fig.~\ref{fig2} 
and ~\ref{fig4}) reveals a very likely $\delta$~Sct pulsation with frequencies in the 
range 30--40\,d$^{-1}$ superimposed over irregular variations on a longer time scale. 
This kind of photometric behaviour is typical of an Herbig Ae star and is caused 
by the partial obscuration of the star's photosphere by circumstellar 
material (see Section~\ref{evolutionary}).

We identify EPIC\,204054556 as a $\gamma$~Dor (GDOR) variable.  EPIC\,204492384 could be 
GDOR, but it looks more like a rotational variable (broad multiple peak at 
 $\sim$1.14\,d$^{-1}$ and harmonic at $\sim$2.28\,d$^{-1}$).  The multiple close frequencies 
at $\sim$ 1.14\,d$^{-1}$ can be attributed to differential rotation. Similarly, 
EPIC\,205002311 might be a rotational variable because the harmonic is 
present, but this classification is uncertain due to the low amplitude and 
the significant noise in this part of the spectrum. 

EPIC\,204506777 (HD\,144548) is a very interesting star.  It is an eclipsing 
binary of Algol type  \citep{Kiraga2012} with period $P = 1.62780$\,d and 
$v \sin i = 80\pm5$\,km\,s$^{-1}$.  There are two small amplitude
eclipses. From the periodogram we find a main period $P = 1.524(1)$\,d.  The light curve at this 
period is roughly sinusoidal and could be interpreted as the rotation period.  
There also seems to be a short regular eclipses at $P = 1.628(1)$\,d
\citep[the same period reported by][see also Fig.\,\ref{fig2}]{Kiraga2012}.  On top of this are deeper eclipses at JD\,2064.32 and 
JD\,2124.01 and a secondary {\it double} eclipse at JD\,2089.94 and JD\,2090.30.  
It is not easy to interpret this variability, but possibly what we are seeing is a 
third body 
which eclipses a close binary with period 
1.628(1)\,d.  The double eclipse is of equal depth, indicating that the close binary 
has equal components.  This is a very interesting system which deserves
further study.


\section{Spectroscopic follow-up}
\label{spectroscopy}

In the previous section we identified several interesting stars worth of
further analysis, especially the seven pulsating stars and the possible
triple system (EPIC\,20450677). To fully exploit the potentialities of the {\it K2}
photometry, especially for the pulsating stars, it is important to
derive high-resolution spectroscopic estimates of the stellar parameters
and of the rotational velocities $v \sin i$. 
To this aim we searched the literature and the available web archives
for high-resolution spectra of our targets. We found data for two stars.

\begin{itemize}

\item
EPIC\,204399980 (HIP\,79476 = HD\,145718) has been studied spectroscopically in 
some detail by \citet{Carmona2010}.  The spectrum displays an inverse P~Cygni 
H$\alpha$ profile, presumably a result of accretion, confirming the Herbig 
Ae nature of this star. They obtained a projected rotational velocity, 
$v \sin i = 100 \pm 10$\,km\,s$^{-1}$ and a centre-of-mass velocity
$v_r = 0 \pm 3$\,km\,s$^{-1}$.

\item
A spectrum of EPIC\,204506777 (HIP\,78977 = HD\,146897), obtained in 2006, 20 June with the 
{\tt FEROS} echelle spectrograph on the 2.2-m MPG telescope (resolution of 48\,000,
exposure time 90\,s) was found in the ESO archive.

\end{itemize}

Nothing was found for the remaining 6 pulsating stars. Hence we decided to
 to observe all of them using the new {\it Catania
  Astrophysical Observatory Spectropolarimeter} (CAOS) 
facility. 

The next section describes these new spectroscopic observations and data 
analysis for the seven observed stars. (see Table~\ref{param} for a list).  We note that 
both EPIC\,204399980 and EPIC\,204506777 were re-observed with CAOS in order to
search for duplicity in the former star and to compare the spectroscopic 
results for the latter.  We also observed the two stars with SC data
which do not pulsate, i.e. EPIC\,204662993 and EPIC\,204926239.
Unfortunately, the weather conditions did not allow observations
of the pulsating stars EPIC 203931628, 204054556 and 204175508.

\subsection{CAOS spectra}

CAOS is a fiber fed, high-resolution spectrograph recently installed at the
Cassegrain focus of the 91-cm telescope of the ``M. G. Fracastoro'' observing 
station of the Catania Astrophysical Observatory (Mt. Etna, Italy).  The 
spectra were obtained in 2015, June and July.  Because the targets were very 
low above the horizon, they could only be observed for about one hour.
We used an exposure time of 3600\,s for all observations. The typical
signal-to-noise ratio of the spectra is in the range 40--80 and the resolution
$R = 45000$  as measured from the ThAr and telluric lines. For details on the 
spectrograph and the data reduction procedure, see \citet{Catanzaro2015}. 

The spectrum synthesis approach followed in this paper to derive
  fundamental stellar parameters is based on the unix port of the
  ATLAS9-SYNTHE suite of codes
  \citep[][]{kur81,kur93,kur93b,sbordone04}. This approach has been
  successfully used in a number of papers in the recent literature
  \citep[see e.g.][just to quote some recent work]{niemc15,catanzaro14,hubrig14,doyle14,paunzen14,Catanzaro2015}.
  The website by Fiorella
  Castelli\footnote{http://www.oact.inaf.it/castelli/} contains a
  description of all the improvements of the Kurucz codes and related
  updates in the input physics performed in the latest years. A
  comparison between SYNTHE \citep{kur81} and SYNTH \citep{valenti96}
  was already performed by \citet{catanzaro13}, who demonstrated that
  the results provided by the two codes are perfectly consistent. We also note
  that the new version of ATLAS, i.e. ATLAS12 \citep{Kurucz1997}, differs from the
  previous ATLAS9 only in the method adopted to compute the opacity.
  A comparison between ATLAS9 and ATLAS12 synthesis was performed by
  e.g. \citet{catanzaro12}, who found that the  two codes give consistent results 
if the metallicity of the target star is not far from the adopted solar one.

\begin{figure*}
\includegraphics[width=16cm]{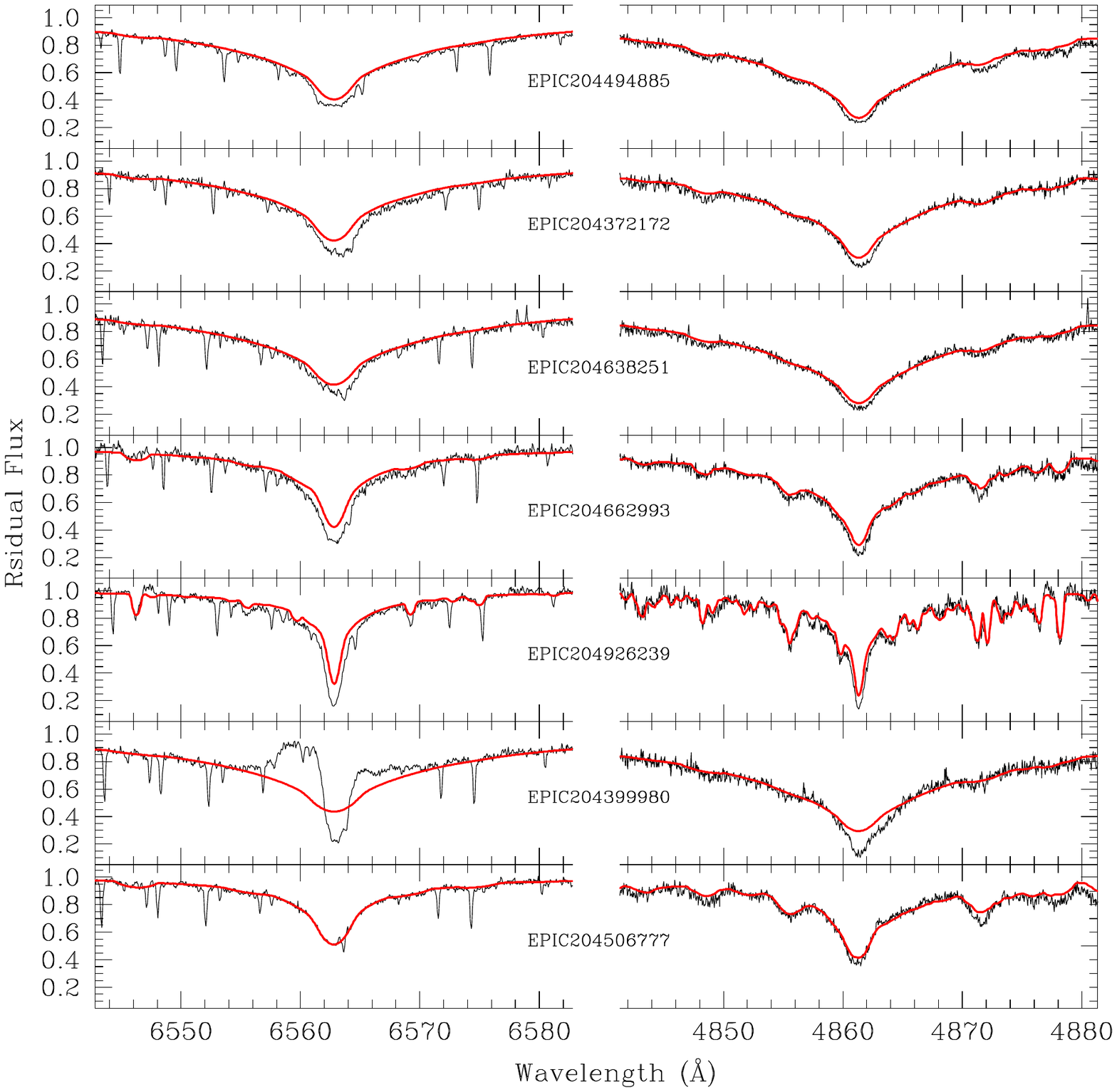}
\caption{Portions of the spectral echelle orders centered around H$\alpha$ and H$\beta$ for the program stars. 
          Superposed (red lines) the synthetic spectra computed with the parameters reported in Table~\ref{param}.}
\label{balmer}
\end{figure*}

\begin{figure}
\centering 
\includegraphics[bb=18 144 350 718 , width=8.5cm]{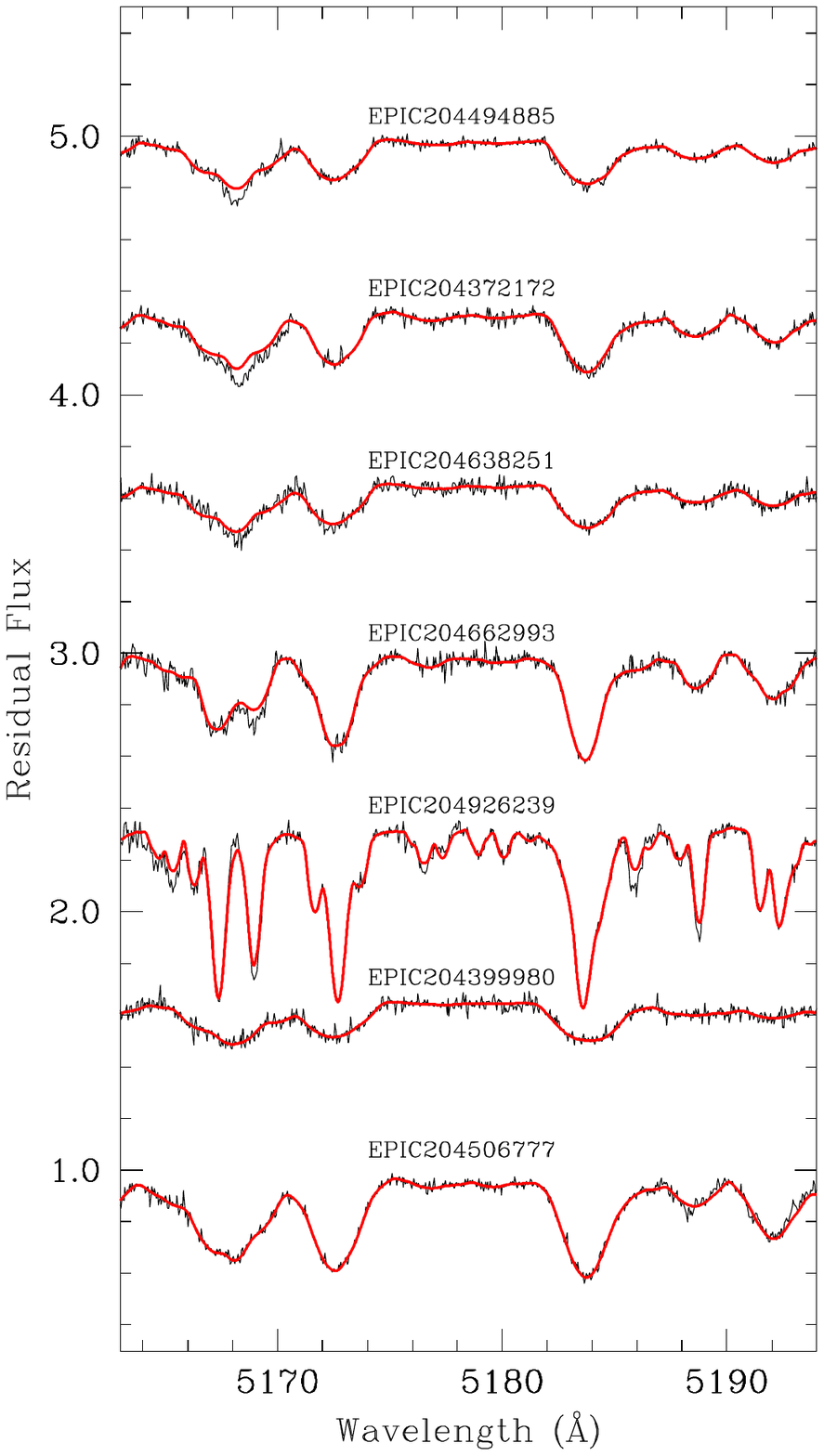}
\caption{Spectral synthesis in the range of the Mg{\sc i} triplet 5150 - 5200 {\AA}}
\label{mag}
\end{figure}

In this paper, since the targets likely have 
circumstellar material that could partially fill the H$\alpha$ line, we have
derived the effective temperatures minimizing the difference
between observed and synthetic H$\beta$ profiles, using as
goodness-of-fit parameter the $\chi^2$ defined as

\begin{equation}
\chi^2\,=\,\frac{1}{N}\sum \left(\frac{I_{obs} - I_{th}}{\delta 
    I_{obs}}\right) 
\end{equation}

\noindent 
where N is the total number of points, I$_{obs}$ and I$_{th}$ are the
intensities of the observed and computed profiles, respectively, and
$\delta I_{obs}$ is the photon noise. Synthetic spectra were generated
in three steps: i) we computed LTE atmospheric models using the ATLAS9
code; ii) the stellar spectra were synthesized
using SYNTHE; iii) the spectra were convolved for the
instrumental and rotational broadening\footnote{Only for HIP\,79476 
this method has not been applied because of the strong emission in the
H$\alpha$ profile and the possible filling of the H$\beta$. In
this case we use the ionization equilibrium of iron to estimate the effective
temperature}.
We used as initial guesses for
temperatures the values derived by \citet{Pecaut2012}, that we
report for the sake of clarity in Table~\ref{param}. In
Fig.~\ref{balmer} we plot the synthetic profiles (even for
H$\alpha$) superposed on the observed spectra. Moreover, we computed
the $\log g$ and the $v \sin i$ by spectral synthesis of the region
around Mg{\sc i} triplet at $\lambda \lambda$5167-5183 {\AA} (see
Fig.~\ref{mag}). The resulting values are listed in Table~\ref{param}.

\begin{figure}
\includegraphics[width=8.0cm,angle=0]{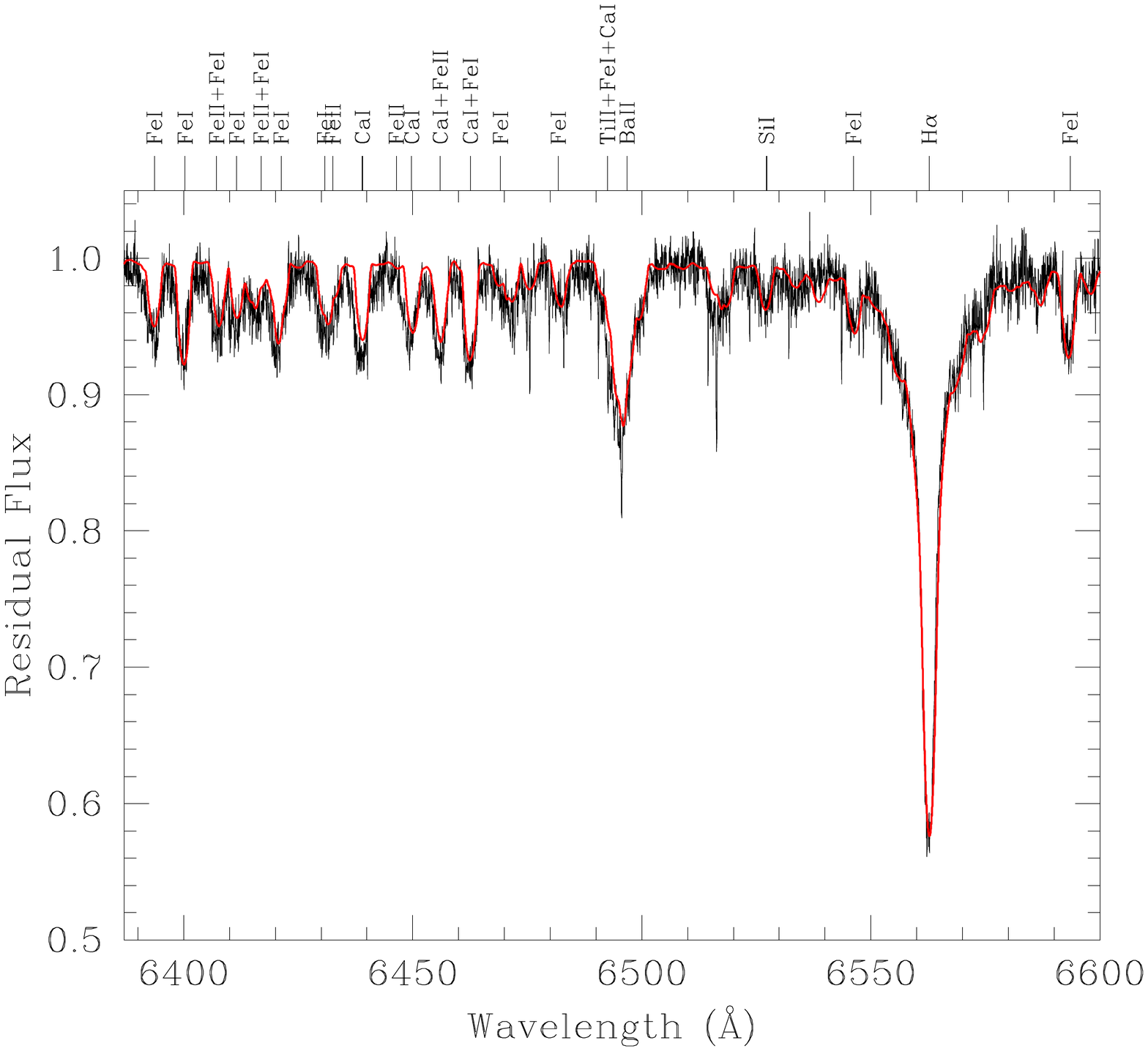}
\caption{Portion of the spectral echelle order in the region of H$\alpha$ for HIP\,78977. Superposed 
(red line) the synthetic spectra computed with the parameters reported in Table~\ref{param}.}
\label{figSpec}
\end{figure}

In a completely similar way we analysed the FEROS spectrum for
EPIC\,204506777 (see Fig.~\ref{figSpec}), which was obtained in better conditions with
respect to the CAOS one (USco is high in sky in the Southern hemisphere).
As a result we found a value of the radial velocity of
$V_{rad}=-13.75\,\pm$\,0.49 
km s$^{-1}$ (JD\,=\,2453906.6783).  Comparison
with the value reported in Table~\ref{param} reveals a clear variation
due to the motion of the binary or triple system. Further data will be
obtained in the future to build up a complete radial velocity curve.

An inspection of Table~\ref{param} reveals that there is an overall
good agreement within the errors between literature \citep[][and
references therein]{Pecaut2012} and CAOS effective temperatures. The
only relevant exceptions are EPIC\,204399980 (HIP\,79476) and 204494885 (HIP\,80130). 
For the first star we find an effective temperature about 500 K larger than
that reported by \citet{Pecaut2012}, who in turn refer to
\citet{Vieira2003}. However, on the basis of
high-resolution spectroscopy, \citet{Carmona2010} discussed the spectral type of this star,
concluding that it cannot be later than A5Ve and estimated a  T$_{\rm eff}^{spec}$=8200
K, almost identical to our value. On this basis we believe that our
estimate of T$_{\rm eff}^{spec}$=8100$\pm$300 K is sound. In passing, we note that our CAOS spectrum allows
us to evaluate the possible multiplicity of EPIC\,204399980, by
comparing our determination of V$_{rad}$=$-$7.4$\pm$6.4 km s$^{-1}$ with
\citet{Carmona2010}'s  V$_{rad}$=0$\pm$3 km s$^{-1}$. The two values
agree within $\sim$ 1 $\sigma$, hence we conclude
that at least  with present data, there is no evidence of
multiplicity for this star.

As for  EPIC\,204494885, the previous effective temperature evaluation dates
back to \citet{Houk1988} who classified the stars visually on
objective-prism plates. Therefore their effective temperature
estimate cannot be accurate. Also in this case we decided to take our value.

\section{Evolutionary status of the investigated stars.}
\label{evolutionary}

In the context of the present paper it is important to establish the
evolutionary status of the target stars, particularly those pulsating
as $\delta$ Sct or $\gamma$ Dor.  Figure~\ref{fig5} shows the HR
diagram for the A- and F-type stars members of the USco association
according to analysis of \citet{Pecaut2012}. For the stellar
parameters we have adopted the values listed in columns (9) and (10)
of Table~\ref{tabGen}, except for the four $\delta$ Sct stars for whom
we used our spectroscopically determined values. As we discussed in
Sect.~\ref{spectroscopy}, there is some disagreement between our own
results and those of \citet{Pecaut2012}. However, we note that the use
of \citet{Pecaut2012} data would not affect significantly our
conclusions.

The HR diagram in Figure~\ref{fig5}, includes some
selected evolutionary tracks and isochrones from the models by
\citet{Tognelli2011}. Note that \citet{Pecaut2012} used for their
analysis a combination of empirical isochrones and the models of
\citet{Dotter2008} and found an age for USco of 11$\pm$1 Myr. As seen
in Fig.~\ref{fig5}, we confirm this estimate since the position of
almost all the A-type stars is compatible within the errors with the
10~Myr isochrone of the models of \citet{Tognelli2011}.

So far, we have assumed that all the stars investigated in this paper
are in the PMS phase.  One can argue that their membership to the USco
association should guarantee their youth, considering that not only
\citet{Pecaut2012}, but also previous studies report an age of 5-10
Myrs \citep[see, e.g.][and references
therein]{deZeeuw1999,Luhman2012} for this association. However, as a check, we now proceed
to provide further evidence on this assumption.

In general, the distinctive properties of young A and F-type stars
(also known as Herbig Ae stars) are the presence of H$\alpha$ emission
(or filling of the line) in their spectrum, a luminosity class III--V,
and the presence of excess infrared emission due to circumstellar dust
\citep[see e.g.][and reference therein]{vandenancker1998}.  Now, all
the stars of our sample satisfy the criterion on the luminosity
class. As for the presence of H$\alpha$ emission, only the spectrum of
EPIC\,204399980 shows it unambiguously (cf.  Sect.~\ref{spectroscopy}
and below), whereas for the remaining stars a partial filling of the
H$\alpha$ line cannot be excluded.

The disk population of USco has been thoroughly investigated by
\citet{Luhman2012}. Their results for the stars in common with our
sample are shown in column (11) of Table~\ref{tabGen}. About half of
our stars show the presence of a disk, even if in most cases it is an
evolved (debris, transitional) structure, possibly devoid of gas.
The only exception is EPIC\,204399980 with its clear signature of
accretion in the H$\alpha$ profile. Not surprisingly, its light curve
displays long term, irregular photometric variations due to the
variable obscuration of the photosphere produced by circumstellar
dust, as commonly found among Herbig Ae stars \citep[see e.g.][and
references therein]{Herbst1994}.  An obvious interpretation would be
that the $\delta$ Sct star EPIC\,204399980 is one of the least evolved
stars of our sample, much younger that the rest.  However, the
observed position in the HR diagram very close to the ZAMS runs
contrary to this hypothesis, unless the stellar luminosity has been
severely underestimated due to a wrong extinction correction.

Considering the other pulsating stars, according to the results of
\citet[][]{Luhman2012}, EPIC\, 203931628, 204054556, and 204372172
have a debris/evolved transitional disk, a clear indication of their
PMS nature.  In the case of EPIC\, 204175508, 204494885, and 204638251
no infrared excess was found in the literature and thus no evidence
for the presence of a circumstellar disk. However, the absence of a
significant infrared excess does not necessarily mean that these stars
are evolved objects, not belonging to the USco association.  Recent
studies based both on Spitzer photometry up to 70 $\rm \mu m$
\citep[][]{Carpenter2006,Carpenter2009} and spectroscopy
\citep[][]{Dahm2009} have shown that in general the disks in USco are
more evolved than those observed in younger star forming regions. Only
$\sim$20\% of the stars have an infrared excess at any wavelength,
mostly due to debris rather than primordial disks.  On the basis of
these considerations, unless the membership evaluation by
\citet[][]{Pecaut2012} is wrong, we can consider all the new pulsators
as young intermediate-mass stars with an age of $\sim$10~Myr.


\begin{table*}
\centering 
\caption{Results obtained from the spectroscopic analysis of the 
  sample of  stars presented in this work. The different columns 
  show: identification, literature \citep{Pecaut2012} and spectroscopic effective temperatures,
  surface gravities ($\log g$), rotational velocities (v\,$\sin i$), heliocentric julian day of the 
  observations and radial velocities. 
}
\begin{tabular}{crcccccr}
\hline 
\hline 
\noalign{\medskip}
EPIC &  HIP~  & T$_{\rm eff}^{Lit}$& T$_{\rm eff}^{spec}$ &  $\log g$ &   v $\sin i$ ~~  & JD          &  V$_{rad}$~~  \\
     &        &           (K)      &           (K)        &           &  (km s$^{-1}$)~~~& (2457100.+) & (km s$^{-1}$) \\
\noalign{\medskip}
\hline 
\noalign{\smallskip}
204506777 & 78977 & 6150\,$\pm$\,150 & 6300\,$\pm$\,200 & 4.10\,$\pm$\,0.15 &  80\,$\pm$\,5  & 78.4706 &            15.6   \,$\pm$\,     3.0    \\
204399980 & 79476 & 7500\,$\pm$\,200 & 8100\,$\pm$\,300 & 4.00\,$\pm$\,0.15 & 110\,$\pm$\,5  & 79.4551&        $-$7.4     \,$\pm$\,   6.4    \\
204926239 & 79606 & 6150\,$\pm$\,150 & 6100\,$\pm$\,150 & 3.80\,$\pm$\,0.15 &  22\,$\pm$\,2  & 80.4255 &     $-$24.1      \,$\pm$\,  1.0     \\
204662993 & 79977 & 6700\,$\pm$\,100 & 6500\,$\pm$\,150 & 4.00\,$\pm$\,0.15 &  55\,$\pm$\,3  & 85.4380 &        $-$6.2     \,$\pm$\,   2.7    \\
204638251 & 80059 & 7800\,$\pm$\,250 & 8000\,$\pm$\,250 & 3.95\,$\pm$\,0.15 &  90\,$\pm$\,5  & 97.3430 &              7.1  \,$\pm$\,      5.7   \\
204372172 & 80088 & 7500\,$\pm$\,150 & 7600\,$\pm$\,200 & 3.80\,$\pm$\,0.15 &  80\,$\pm$\,5  & 96.4365 &        $-$9.1     \,$\pm$\,   6.0    \\
204494885 & 80130 & 7500\,$\pm$\,150 & 8000\,$\pm$\,250 & 3.80\,$\pm$\,0.15 &  85\,$\pm$\,5  & 78.4212 &      $-$51.9      \,$\pm$\,  6.0    \\
\noalign{\smallskip}
\hline 
\end{tabular}
\label{param}
\end{table*}

\section{The pulsating stars}

Recently, \citet{Zwintz2014} have observed a large sample of PMS
 $\delta$~Sct stars and determined their effective
temperatures and surface gravities from spectroscopic data.  An
important result of this study is that the highest observed $p$-mode
frequency, $\nu_{\rm max}$, correlates quite well with the location of
the star in the HR diagram since it scales as $gT_{\rm
  eff}^{-1/2} \sim M T_{\rm  eff}^{3.5} L^{-1}$ \citep[see][for details]{Brown1991,Kjeldsen1995,Zwintz2014}. 
The coolest and least-evolved stars (i.e.
closest to the birth-line) have the lowest values of $\nu_{\rm max}$,
while the hottest and most evolved stars (closest to the ZAMS) have
the highest values of $\nu_{\rm max}$.  \citet{Zwintz2014} interpret
the observed trend as due to the behaviour of the acoustic cut-off
frequency. The latter is directly proportional to the square root of
the star's mean density and therefore it should increase as a star
contracts and its density rises. Of course, this property can only be
observed in young stars whose radius varies with age. Indeed, for the
more evolved, core hydrogen-burning $\delta$~Sct stars there is no
correlation of $\nu_{\rm max}$ with the stellar parameters.  Using the
observed trend in a homogenous sample of pulsating stars in the young
cluster NGC~2264, \citet{Zwintz2014} were able to quantify the age
interval during which star formation took place in the parent
molecular cloud at the level of at least 5~Myr.

As discussed in Sect.~\ref{Classification}, we have identified six new
pulsating stars: namely, EPIC\,203931628, 204175508, 204372172,
204399980, 204494885 and 204638251.  Unfortunately, since the Nyquist
frequency limit of our data is about 24\,d$^{-1}$, the frequencies
extracted from the LC mode data are subject to ambiguity preventing an
accurate determination of $\nu_{\rm max}$.  On the other hand, four
stars have been observed in SC mode and in this case we could
calculate the values of $\nu_{\rm max}$. It turns out that in all
cases $\nu_{\rm max}\sim 42\,$d$^{-1}$, apart from EPIC\,204372172
with $\nu_{\rm max} \approx 33$\,d$^{-1}$. Furthermore, all these
stars have similar properties in terms of effective temperatures and
luminosities. Therefore, since the relationship between $\nu_{\rm max}$ 
and the stellar parameters is approximate, we can only draw
some qualitative conclusion from the knowledge of the cutoff
frequencies.

First, the similar values of $\nu_{\rm max}$ could mean that there is
little or no age spread in USco among the late A-type stars. Even if
EPIC\,204372172 has a smaller $\nu_{\rm max}$, the position in the HR
diagram is consistent with that of the other stars, preventing us from
finding any significant difference in age from the comparison with the
isochrones. Second, the fact that EPIC\,204399980, the star that we
consider to be the youngest of our sample on the basis of the presence
of a fully developed circumstellar disk, shows a value of $\nu_{\rm
  max}$ as high as that of the other $\delta$ Sct stars may appear
difficult to interpret.  A possibility is that the relation $\nu_{\rm
  max}$--age found by \citet{Zwintz2014} is not always present in
young clusters/associations \citep[][]{Stahler2014}. Vice versa, and
more likely, all the four pulsators of USco have approximately the
same age, but EPIC\,204399980 somehow managed to maintain its disk for
a longer time. This is not an unusual situation considering that
actively accreting stars older than 6-8~Myr and still surrounded by
circumstellar disks have been found in populous clusters, such as
NGC~6611 \citep[][]{DeMarchi2013}.

In any case, this result is different from that observed in  NGC\,2264 studied by
\citep[][]{Zwintz2014}. Interestingly, NGC~2264 is known to have a
median age of $\sim$3 Myr, and a significant dispersion of the order
of $\sim$~5 Myr \citep[see also][]{Dahm2008}. In the case of USco, the
age difference of its members seems to be very small. However, looking
at their distribution in the HR diagram shown in Fig.~8, we notice
that there are several stars of USco that fall within the instability
strip and that lie close to the 5~Myr isochrone (see the light green
symbols). Thus, one can argue for an indication of a possible spread
in isochronal ages in USco members of the same magnitude of that more
solidly derived in NGC~2264. Unfortunately, these putative younger
stars have not been observed by {\it Kepler}, but they are ideal
targets for future dedicated observations to search for pulsation and
therefore to probe the presence of a possible age spread in this
association.

\begin{table}  
\begin{center}  
\caption{List of pulsation frequencies and amplitudes for the four
candidate PMS $\delta$~Sct stars observed in SC mode. EPIC\,204054556 is 
a candidate $\gamma$~Dor star.  Given the problematic light curve
at low frequencies, for star EPIC\,204399980, only frequencies larger 
than 10\,d$^{-1}$ are listed. Errors on each value of frequency and amplitude are
shown within parenthesis.}
\label{freqs}
\vspace{2mm}  
\begin{tabular}{rrrrrr}
\hline  
\multicolumn{1}{c}{$n$}             &
\multicolumn{1}{c}{$\nu_n$}         &
\multicolumn{1}{c}{$A_n$}           &
\multicolumn{1}{c}{$n$}             &
\multicolumn{1}{c}{$\nu_n$}         &
\multicolumn{1}{c}{$A_n$}           \\
\multicolumn{1}{c}{}                &
\multicolumn{1}{c}{d$^{-1}$}        &
\multicolumn{1}{c}{ppt}             &
\multicolumn{1}{c}{}                &
\multicolumn{1}{c}{d$^{-1}$}        &
\multicolumn{1}{c}{ppt}            \\
\hline 
\multicolumn{3}{c}{\bf 204638251} & \multicolumn{3}{c}{\bf 204494885} \\
  1 & 26.3734(1)  & 4.09(1) &  1 & 22.3371(1) & 0.741(2) \\
  2 & 22.2123(1)  & 3.20(1) &  2 & 28.6862(1) & 0.445(2) \\
  3 & 20.4388(1)  & 2.99(1) &  3 & 26.1806(1) & 0.395(2) \\
  4 & 21.2569(1)  & 2.57(1) &  4 & 23.1134(1) & 0.332(2) \\
  5 & 32.2632(1)  & 2.57(1) &  5 & 20.2332(1) & 0.277(2) \\
  6 & 27.8936(1)  & 2.46(1) &  6 & 32.1927(1) & 0.216(2) \\
  7 & 34.8966(1)  & 2.28(1) &  7 & 19.0010(1) & 0.191(2) \\
  8 & 20.5055(1)  & 2.28(1) &  9 & 28.6459(1) & 0.150(2) \\
  9 & 26.8491(1)  & 1.83(1) & 10 & 21.9472(2) & 0.102(2) \\
 10 & 29.8055(1)  & 1.59(1) & 11 & 27.7289(2) & 0.088(2) \\
 11 & 23.1606(1)  & 1.50(1) & 12 & 18.9506(2) & 0.081(2) \\
 12 & 19.4065(1)  & 1.45(1) & 13 & 29.7855(2) & 0.079(2) \\
 13 & 25.4210(1)  & 1.43(1) & 14 & 20.3818(2) & 0.079(2) \\
 14 &2 4.2454(1)  & 1.38(1) & 15 & 31.4909(2) & 0.070(2) \\     
 15 & 32.0598(1)  & 1.11(1) & 16 & 35.8862(3) & 0.055(2) \\ 
 16 & 20.8422(1)  & 1.06(1) & 17 & 24.6552(3) & 0.050(2) \\ 
 17 & 17.5276(1)  & 1.04(1) & 18 & 41.5304(3) & 0.046(2) \\ 
 18 & 27.4845(1)  & 1.00(1) & 19 & 17.8007(3) & 0.045(2) \\ 
 19 & 32.6900(1)  & 0.99(1) & 20 & 28.9853(4) & 0.044(2) \\
 20 & 28.0091(1)  & 0.89(1) & 21 & 35.2287(4) & 0.041(2) \\
 21 & 34.3108(1)  & 0.89(1) & 22 & 23.5385(4) & 0.037(2) \\
 22 & 19.6051(1)  & 0.88(1) & 23 & 25.4051(4) & 0.036(2) \\
 23 & 37.9624(1)  & 0.87(1) & 24 & 29.9793(5) & 0.034(2) \\
 24 & 18.6522(1)  & 0.81(1) & 25 & 40.9774(5) & 0.029(2) \\
 25 & 19.7238(1)  & 0.77(1) & 26 & 31.1484(5) & 0.029(2) \\
 26 & 24.8841(1)  & 0.73(1) & 27 & 29.4519(5) & 0.029(2) \\
 27 & 29.1318(1)  & 0.68(1) & 28 & 31.0579(6) & 0.027(2) \\
 28 & 28.1713(2)  & 0.58(1) & 29 & 29.6588(6) & 0.027(2) \\   
 29 & 21.9698(2)  & 0.58(1) &    &            &          \\                                 
 30 & 41.0177(2)  & 0.56(1) &\multicolumn{3}  {c}{\bf 204054556}\\            
 31 & 24.8410(2)  & 0.54(1) &  1 &  0.9547(4) & 0.94(3)  \\                                
 32 & 27.4063(2)  & 0.52(1) &  2 &  1.3645(4) & 0.84(3)  \\                                
 33 & 38.0033(2)  & 0.52(1) &  3 &  0.9129(4) & 0.91(3)  \\
 34 & 23.2401(2)  & 0.48(1) &  4 &  0.9907(4) & 0.82(3)  \\
 35 & 26.1315(2)  & 0.47(1) &  5 &  1.3433(4) & 0.73(3)  \\
 36 & 26.5991(2)  & 0.45(1) &  6 &  0.9743(5) & 0.46(3)  \\
 37 & 20.7439(2)  & 0.45(1) &  7 &  0.3096(5) & 0.50(3)  \\
 38 & 22.2047(2)  & 0.39(1) &  8 &  0.9346(5) & 0.53(3)  \\
    &             &         &    &            &          \\
\multicolumn{3}{c}{\bf 204372172} &  \multicolumn{3}{c}{\bf 204399980}\\
  1 & 27.1136(1) & 1.073(5) &  1 & 38.5484(1) & 9.11(2)  \\
  2 & 25.7598(1) & 0.869(5) &  2 & 36.1364(1) & 6.39(2)  \\
  3 & 25.9414(1) & 0.819(5) &  3 & 35.9815(1) & 6.09(2)  \\
  4 & 32.9888(1) & 0.770(5) &  4 & 33.7720(2) & 2.51(2)  \\
  5 & 28.2886(1) & 0.520(5) &  5 & 28.7870(2) & 2.28(2)  \\
  6 & 24.5972(1) & 0.315(5) &  6 & 44.4148(2) & 1.98(2)  \\
  7 & 26.9462(1) & 0.311(5) &  7 & 31.1505(2) & 1.82(2)  \\
  8 & 29.4115(2) & 0.227(5) &  9 & 36.7527(2) & 1.72(2)  \\
  9 & 24.3411(2) & 0.156(5) & 10 & 21.1801(2) & 1.69(2)  \\
 10 & 31.0072(3) & 0.119(5) & 11 & 32.5195(2) & 1.26(2)  \\
 11 & 26.2301(4) & 0.098(5) & 12 & 28.8649(3) & 1.06(2)  \\
 12 & 20.5083(5) & 0.066(5) & 13 & 28.8440(3) & 0.79(2)  \\
    &            &          & 14 & 27.5603(3) & 0.68(2)  \\
    &            &          & 15 & 25.6695(4) & 0.55(2)  \\
    &            &          & 16 & 23.4758(4) & 0.46(2)  \\
    &            &          & 17 & 17.3511(5) & 0.38(2)  \\
    &            &          & 18 & 38.7263(5) & 0.38(2)  \\
    &            &          & 19 & 36.1503(5) & 0.38(2)  \\
    &            &          &    &            &          \\
 \hline
 \end{tabular}
 \end{center}
 \end{table}


\begin{figure}
\includegraphics[width=10.5cm]{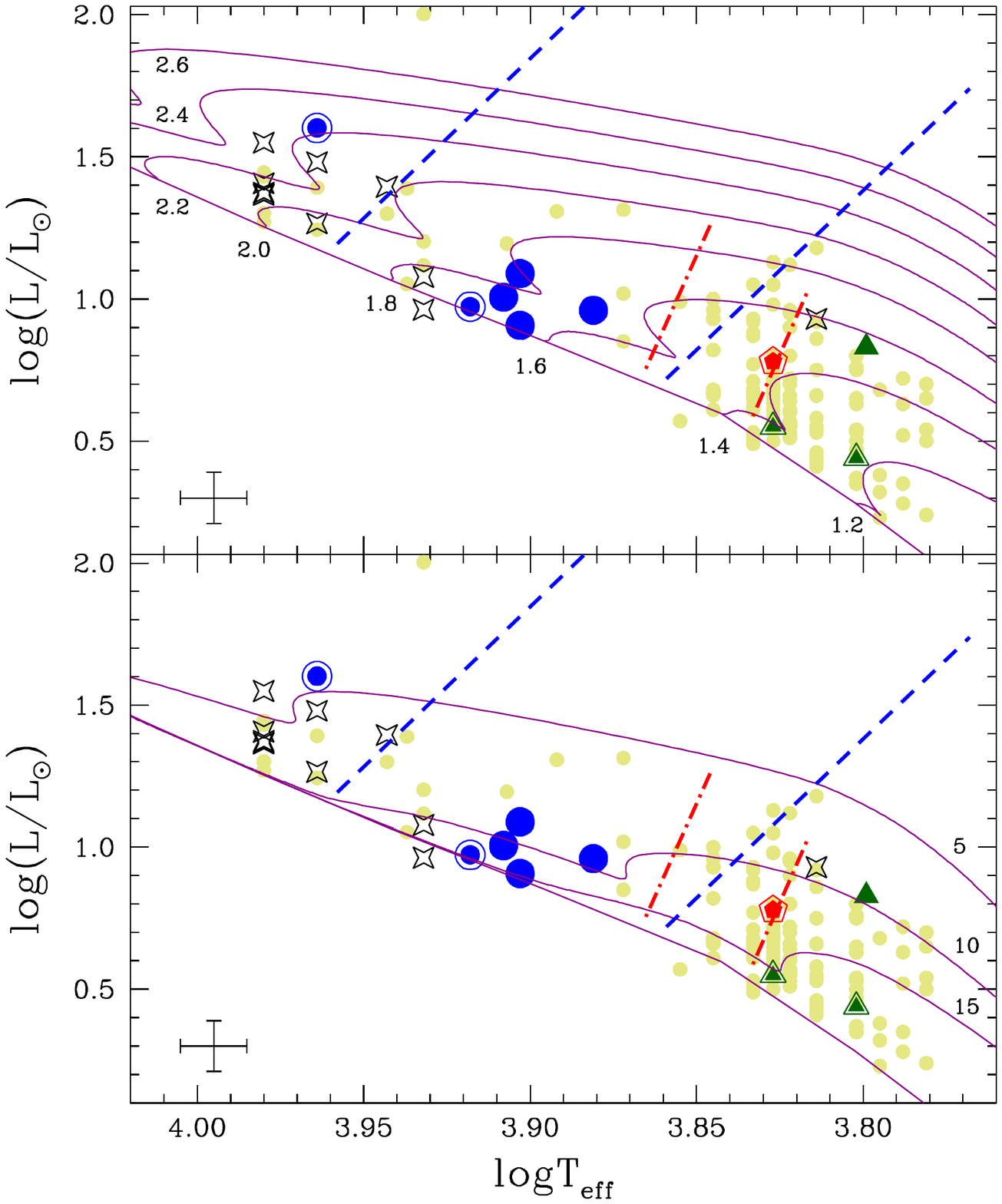}
\caption{HR diagram for the A-F stars members of the USco
  association according to \citet{Pecaut2012} except for the four
  $\delta$ Sct observed with CAOS. For these stars we used the results
  of Table~\ref{param}. Light green symbols show objects not observed
  by {\it Kepler}. Black four-starred symbols represent constant
  stars. Green triangles, red pentagons and blue circles show
  binary/rotational, $\gamma$ Dor variables and $\delta$ Sct,
  respectively. For these variables, empty-filled and filled symbols
  represent data possessing only LC and LC-SC cadences, respectively.
  For comparison purposes we overplot the $\delta$\,Sct (blue dashed
  lines) and the theoretical edges of the $\gamma$\,Dor (red
  dotted-dashed lines) instability strips by \citet{breger98} and
  \citet{guzik}, respectively. We note that the mixing-length
  parameter adopted in the computation of $\gamma$\,Dor instability
  strip was 1.87.  The solid lines in the top panel show the ZAMS and
  selected evolutionary tracks from \citet{Tognelli2011}. Similarly,
  in the bottom panel solid lines show isochrones for 5,10,and 15 Myrs
  as taken from the same Authors.  The evolutionary models and
  isochrones were calculated for a solar mixture with Z = 0.015, Y = 0.278 and
  mixing-length = 1.68.}
\label{fig5} 
\end{figure}



It is interesting to note that low frequencies are almost certainly
present in the PMS $\delta$~Sct stars discussed here.
However, one needs to be careful in making this conclusion because the K2 data
are not very reliable at low frequencies due to drift of the stars
across the CCD.  Nevertheless, the low frequencies in these stars are
of sufficiently high amplitude that one can be confident about their
identification.  The presence of low frequencies is particularly
interesting in PMS stars because envelope convection is
much more extended in this phase, particularly at early stages when
the star radius is still expanded.  Therefore, one may expect that the
convective blocking mechanism, responsible for the onset of $g$-modes \citep[][]{Guzik2000},
could be more active among these stars. In turn, the study of these
low frequency modes is fundamental to investigate the evolutionary
status and the internal structure of the intermediate-mass PMS
pulsators \citep[][]{Suran2001}.

\begin{figure}
\centering 
\includegraphics[width=8.5cm]{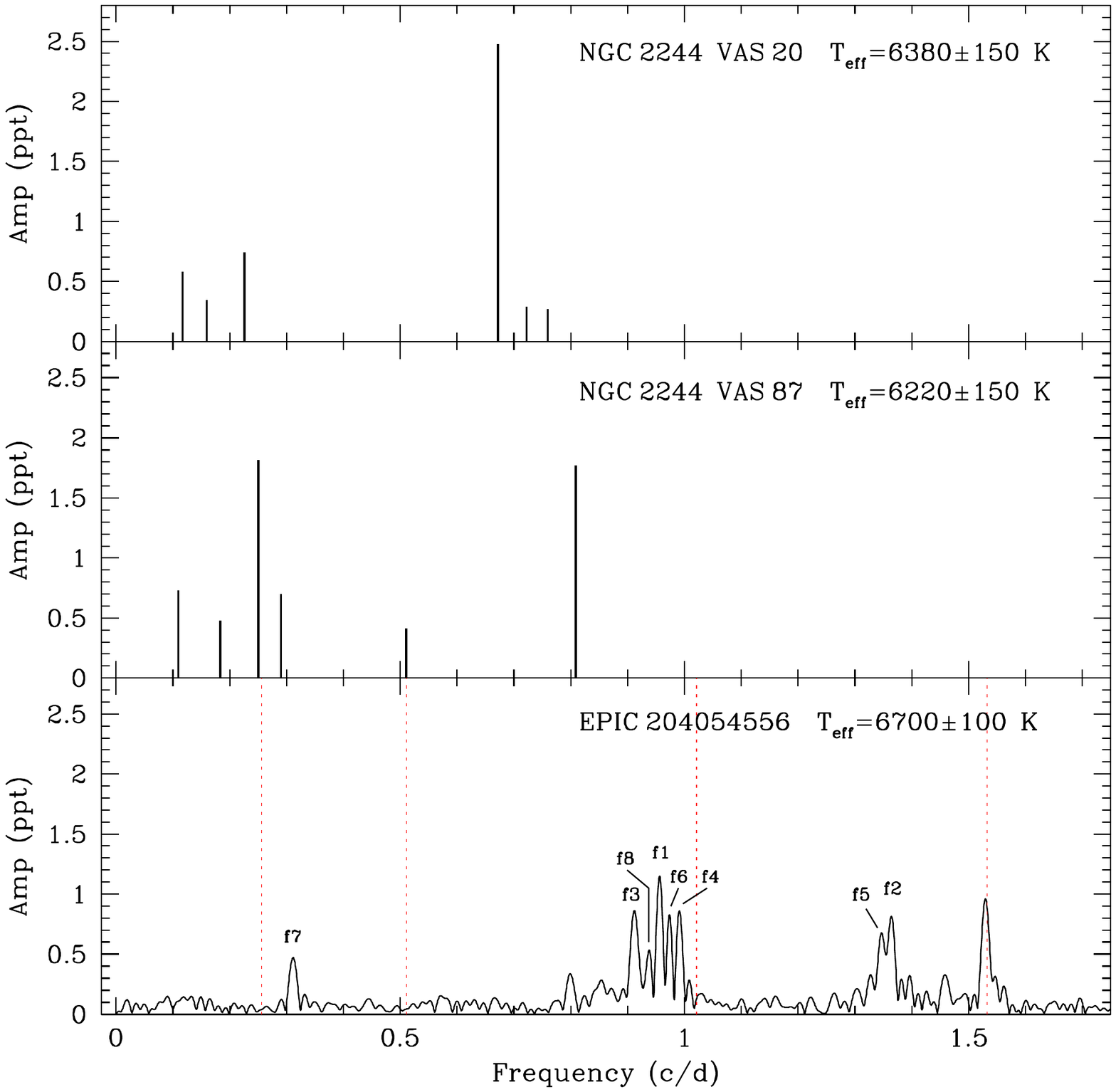}
\caption{Top and medium panels show the schematic periodograms of the PMS $\gamma$~Dor variables VAS\,20 and 
VAS\,87 in NGC\,2244 \citep{Zwintz2013}. Bottom panel displays the
periodogram of the candidate $\gamma$~Dor 
variable in USco, EPIC\,204054556. The labels corresponds to the
frequencies listed in Table~\ref{freqs}. The dashed lines show the
instrumental frequencies.}
\label{figGDor}
\end{figure}

The independent frequencies extracted for the $\gamma$~Dor star 
EPIC\,204054556 are listed in Table~\ref{freqs} and displayed in the
bottom panel of Fig.~\ref{figGDor}. In the same figure, we also
show for comparison the periodgrams of the two other known PMS
$\gamma$~Dor stars, VAS\,20 and VAS\,87 in the young cluster NGC\,2264
\citep{Zwintz2013}.  We see that the frequencies detected in
EPIC\,20405455 are systematically higher than the $\gamma$~Dor stars
of NGC\,2264, a fact that can be attributed to the higher effective
temperature of EPIC\,20405455.  As can be seen in Fig.~\ref{fig5},
this star falls well within the instability strip for $\gamma$~Dor
stars, whereas the two variables in NGC\,2264 are slightly outside
(see Fig.\,9 in \citealt{Zwintz2013}).

As a final consideration, we note that two of the investigated stars,
namely EPIC\, 204239132 and 204242194, do not show any pulsation at a
level of $\sim$0.01 ppt, albeit being placed inside the $\delta$ Sct
instability strip, towards its blue boundary (see~\ref{fig5}). The
effective temperature estimate for these stars is not based on solid
high-resolution spectroscopy \citep[see][and references
therein]{Pecaut2012}. However, even a 500 K error on the effective temperature
cannot place these stars outside the strip. 
The presence of non-variable stars inside the $\delta$ Sct instability strip is a well known long standing
problem \citep[see e.g.][and references therein]{Balona2011a}. A recent discussion of this subject is provided by
\citet{Murphy2015} whose investigation is based on ultra-precise {\it Kepler}  
photometry and high-resolution spectroscopy. Their conclusion is that 
chemically normal\footnote{In A-type chemically peculiar stars, such
  as the Am class, a dumping of the pulsation is expected \citep[see
  e.g.][and references therein]{Balona2011b}.} non-variable stars inside the $\delta$ Sct instability
strip exist but are rare. However, the blue limit of the
instability strip used by these authors \citep[i.e.][]{Dupret2005} is 
significantly redder than that adopted in the present work. 
As a result, according to \citet{Dupret2005} calculations, EPIC\, 204239132 and
204242194 would be close or outside the blue boundary of the $\delta$
Sct instability strip.
In conclusion, given current uncertainties on both effective
temperature measurements and instability strip determinations the
occurrence of  static stars inside the adopted pulsation
region cannot be excluded.

\section{Conclusions}

We have presented the light curves obtained with the {\it Kepler}
spacecraft and the derived periodograms of 27 stars belonging to the
young Upper Scorpius association. This sample contains only A-F type
stars which are known to be subject to pulsation instability of the
$\delta$~Sct and $\gamma$~Dor type during the contraction phase prior
to the arrival on the main sequence.

We have identified six new $\delta$~Sct variables and one $\gamma$~Dor
star (EPIC\,204054556/HIP\,79054) which are most likely PMS stars just
about to arrive on the ZAMS.  Four of the $\delta$ Sct variables,
namely EPIC\,204372172 (HIP\,80088) EPIC\,204399980 (HIP\,79476)
EPIC\,204494885 (HIP\,80130), and EPIC\,204638251 (HIP\,80059) were
also observed in short cadence mode (time sampling 1-min) allowing us to
resolve their entire frequency spectrum. For the remaining two
$\delta$ Sct stars, EPIC\,203931628 (HIP\ 80196) and EPIC\,204175508
(HIP\ 77960), we only got long cadence data (time sampling $\sim$30-min)
that were enough to unambiguously identify their variability. Thanks to
these new data, USco is the second young cluster after NGC\,2264 that
hosts several PMS $\delta$ Sct/$\gamma$ Dor variables discovered through
space observations.

One of the stars observed in short cadence, EPIC\,204506777
(HIP\,78977) appears to be a rotational variable in a hierarchical
triple system.  This is a particularly important system since future
radial velocity measurements will allow us to obtain accurate
determination of the mass of the star, thus setting strong constraints
on its pulsation properties and on evolutionary models.

Follow-up spectroscopic observations of the newly discovered variables
observed in short cadence allowed us to obtain new estimates of
T$_{\rm eff}$, $\log g$, v $\sin i$, and V$_{rad}$. Our second epoch
spectroscopy for EPIC\,204506777 (the first epoch was obtained from
archive data) reveals a significant change of center-of-mass radial
velocity for this star, confirming its multiplicity. However,
additional data are required to construct a complete radial velocity
curve.

We have used the newly determined stellar parameters of four of the
$\delta$ Sct stars to place them in the HR diagram and obtain an
independent estimate on their age, confirming the value of
$\sim$10~Myr for the members of the USco association.  
Interestingly, their
measured $v\sin i$ values are less or close to
$\sim$100\,km\,s$^{-1}$, thus  
the effect of rotation on the pulsation frequencies
should be less extreme and the frequency spectra more suitable for
comparison with stellar pulsation models \citep[e.g.][]{Ripepi2011}.  A
quantitative analysis of the new $\delta$ Sct and $\gamma$~Dor stars
with the purpose to determine their masses and evolutionary stage will
be the subject of a forthcoming paper.

\section*{Acknowledgments} 

We thank our anonymous referees for their valuable comments.

LAB thanks the SAAO and National Research Foundation of South Africa for
financial support.

This work made use of PyKE (Still \& Barclay 2012), a software package
for the reduction and analysis of Kepler data.  This open source
software project is developed and distributed by the NASA Kepler Guest 
Observer Office.  

This research has made use of the SIMBAD database and 
VizieR catalogue access tool, operated at CDS, Strasbourg, France.

The authors gratefully acknowledge the {\it Kepler}
team and the Guest Observer Office whose out--standing efforts 
have made these results possible.

This paper includes data collected by the Kepler mission. Funding for
the Kepler mission is provided by the NASA Science Mission
directorate. 

All of the data presented in this paper were obtained from the
Mikulski Archive for Space Telescopes (MAST). STScI is operated by the
Association of Universities for Research in Astronomy, Inc., under
NASA contract NAS5-26555. Support for MAST for non-HST data is
provided by the NASA Office of Space Science via grant NNX09AF08G and
by other grants and contracts.

\label{lastpage} 
 
\end{document}